\documentclass[11pt]{article}

\usepackage[font=small,labelfont=bf]{caption}

\usepackage{enumitem}
\usepackage{graphicx}
\usepackage{xcolor}
\usepackage{blkarray}
\usepackage{amsmath,amssymb}
\usepackage{bbm}
\usepackage{mathtools}
\usepackage{pdflscape}
\usepackage{float}
\usepackage{rotating}
\usepackage{titlesec}
\usepackage{tcolorbox}
\usepackage{lscape}
\usepackage{rotating}

\usepackage[
citestyle=numeric,
style=numeric,
uniquename=false,
sorting=ynt,
maxbibnames=8
]{biblatex}

\titleformat{\paragraph}
{\normalfont\normalsize\bfseries}{\theparagraph}{1em}{}
\titlespacing*{\paragraph}
{0pt}{3.25ex plus 1ex minus .2ex}{1.5ex plus .2ex}

\DeclareMathOperator{\EX}{\mathbbm{E}}
\allowdisplaybreaks

\begin{document}
\setlength{\abovedisplayskip}{4pt}
\setlength{\belowdisplayskip}{10pt}
\setlength{\abovedisplayshortskip}{4pt}
\setlength{\belowdisplayshortskip}{10pt}

\title{A hybrid transmission model for \textit{Plasmodium vivax} accounting for superinfection, immunity and the hypnozoite reservoir}

\author{Somya Mehra$^{1}$ \and Peter G. Taylor$^1$ \and James M. McCaw$^{1,2,3}$ \and Jennifer A. Flegg$^1$}

\date{%
    $^1$School of Mathematics and Statistics, The University of Melbourne, Parkville,  Australia\\%
    $^2$Centre for Epidemiology and Biostatistics, Melbourne School of Population and Global Health, The University of Melbourne, Parkville, Australia\\
    $^3$Peter Doherty Institute for Infection and Immunity, The Royal Melbourne Hospital and The University of Melbourne, Parkville, Australia\\[2ex]
}

\maketitle

\section*{Abstract}
Malaria is a vector-borne disease that exacts a grave toll in the Global South. The epidemiology of \textit{Plasmodium vivax}, the most geographically expansive agent of human malaria, is characterised by the accrual of a reservoir of dormant parasites known as hypnozoites. Relapses, arising from hypnozoite activation events, comprise the majority of the blood-stage infection burden, with implications for the acquisition of immunity and the distribution of superinfection. Here, we construct a hybrid transmission model for \textit{P. vivax} that concurrently accounts for the accrual of the hypnozoite reservoir, (blood-stage) superinfection and the acquisition of immunity. We begin by analytically characterising within-host dynamics as a function of mosquito-to-human transmission intensity, extending our previous model (comprising an open network of infinite server queues) to capture a discretised immunity level. To model transmission-blocking and antidisease immunity, we allow for geometric decay in the respective probabilities of successful human-to-mosquito transmission and symptomatic blood-stage infection as a function of this immunity level. Under a hybrid approximation --- whereby probabilistic within-host distributions are cast as expected population-level proportions --- we couple host and vector dynamics to recover a deterministic compartmental model in line with Ross-Macdonald theory. We then perform a steady-state analysis for this compartmental model, informed by the (analytic) distributions derived at the within-host level. To characterise transient dynamics, we derive a reduced system of integrodifferential equations (IDEs), likewise informed by our within-host queueing network, allowing us to recover population-level distributions for various quantities of epidemiological interest. In capturing the interplay between hypnozoite accrual, superinfection and acquired immunity --- and providing, to the best of our knowledge, the most complete population-level distributions for a range of epidemiological values --- our model provides insights into important, but poorly understood, epidemiological features of \textit{P. vivax}.

\section{Introduction}
Despite decades of concerted control and elimination efforts, malaria persists as a grave cause of morbidity and mortality in the Global South, yielding an estimated 241 million cases and 627,000 deaths in 2020 alone \parencite{world2021world}. The global malaria burden is largely driven by the parasites \textit{Plasmodium falciparum} and \textit{Plasmodium vivax}, with the transmission of both parasites mediated by \textit{Anopheles} mosquito vectors. In light of its expansive geographical distribution, over three billion people are thought to be at risk of \textit{P. vivax} infection  \parencite{battle2019mapping, battle2021global}.\\

The difficulty of eliminating \textit{P. vivax}, in particular, is compounded by a reservoir of dormant parasites, known as hypnozoites, hidden within the human liver. The consequences of mosquito inoculation for \textit{P. vivax} are two-fold: in addition to causing a primary (blood-stage) infection, an infective bite can lead to the establishment of an (undetectable) batch of hypnozoites in the liver. Following an indeterminate dormancy period, the activation of a single hypnozoite can give rise to (blood-stage) relapse. Long latency periods (in the order of 6 to 9 months) are characteristic of temperate strains, while short latency periods (typically spanning 3 to 6 weeks) tend to be more common in tropical settings \parencite{battle2014geographical, white2016variation}. Hypnozoite activation is believed to be a key driver of superinfection --- which involves the co-circulation of multiple parasite broods in the bloodstream \parencite{popovici2018genomic}.\\

The hypnozoite reservoir also has important implications for the acquisition of immunity \parencite{mueller2013natural}. Relapse-driven exposure to a large number of genetically-distinct clones in early childhood is believed to underpin the dynamics of acquired immunity to \textit{Plasmodium vivax} \parencite{koepfli2013high}. The mechanisms of immune protection for \textit{P. vivax} are multi-faceted and highly complex, but are known to be stage-specific (see \textcite{antonelli2020immunology} for a recent review). The majority of the immune response to \textit{P. vivax} is believed to be targeted towards asexual blood-stage parasites. Here, we distinguish two manifestations of (asexual) blood-stage immunity:
\begin{itemize}
    \item \textit{Clinical/antidisease immunity} reduces the risk or severity of clinical symptoms during a blood-stage infection, with epidemiological data suggesting rapid acquisition \parencite{mueller2013natural, mueller2015development}.
    \item \textit{Antiparasite immunity} modulates the clearance of blood-stage infection \parencite{deroost2016immunological}, and is typically modelled through accelerated parasite clearance rates and/or reduced parasite densities \parencite{griffin2010reducing, white2018mathematical}.
\end{itemize}
\textit{Transmission-blocking immunity}, which modulates the infectivity of sexual blood-stage parasites (gametocytes) to mosquitoes, is also of note, with the mitigation of mosquito-stage development curtailing onward transmission \parencite{gamage1992transmission, mueller2013natural, mueller2015development, de2020immunity}. There is evidence to suggest that the gametocyte circulation is \textit{not} hampered by clinical and antiparasite immunity \parencite{joyner2019humoral}.\\

Other forms of immunity are thought to be of comparatively limited consequence in \textit{natural} transmission settings. Pre-erythrocytic immunity targets sporozoites inoculated through mosquito bites, prior to further liver-stage development. Due to the potential for each sporozoite to develop into a hypnozoite, pre-erythrocytic immune protection has been hypothesised to substantially mitigate the relapse burden \parencite{mueller2013natural, white2017theoretical}; exposure to sporozoites in natural transmission settings, however, is believed to be insufficient to induce strong pre-erythrocytic immune protection \parencite{mueller2013natural}. Likewise, immune responses targeted towards liver-stage parasites, particularly hypnozoites, are poorly understood \parencite{galinski2008plasmodium}, but are generally considered to be relatively minor.\\

The joint dynamics of immunity and the hypnozoite reservoir are of epidemiological interest. The dichotomisation of both hypnozoite carriage and immune status \parencite{kammanee2001basic, ishikawa2003mathematical, aguas2012modeling, roy2013potential} yields, in some senses, the simplest approach for characterising population-level transmission dynamics. Various models of immunity have been proposed under these dichotomised frameworks, ranging from imperviousness to reinfection (until immunity is lost) \parencite{kammanee2001basic}, to an elevated rate of recovery (antiparasite) \parencite{ishikawa2003mathematical}; reduced infectiousness to mosquitoes (transmission-blocking) \parencite{roy2013potential}; and necessarily asymptomatic blood-stage infection (clinical) \parencite{aguas2012modeling}. A slightly extended model of transmission-blocking immunity, superinfection and hypnozoite accrual has been proposed by \textcite{de1991mathematical} --- with the limitation that each individual can harbour up to two broods of hypnozoites and two overlapping relapses, and a discrete immunity level $\{0, 1, 2 \}$.\\

A more comprehensive characterisation of immunity and the hypnozoite reservoir has recently been performed by by \textcite{white2018mathematical}. Under a hypnozoite `batch' model --- whereby hypnozoites are stratified into `batches', each characterised by a constant rate of relapse over the span of an exponentially-distributed lifetime, with an imposed upper bound $K$ on concurrent batch carriage --- \textcite{white2018mathematical} account for the acquisition of both antidisease immunity (which reduces the probability of symptomatic infection) and antiparasite immunity (which results in an elevated rate of parasite clearance and a reduced probability of detection via light miscroscopy). In addition to being restricted to short-latency strains, the framework of \textcite{white2018mathematical} ignores size variation in parasite inocula, as we noted in \textcite{mehra2022hypnozoite}. Further, \textcite{white2018mathematical} do not explicitly account for superinfection.\\

Here, we seek to characterise the interplay between hypnozoite accrual, superinfection and acquired immunity, for both short- and long-latency strains. In \textcite{mehra2022superinf}, we have recently proposed a transmission model for \textit{P. vivax} that explicitly accounts for superinfection and (short-latency) hypnozoite accrual; our model can be viewed as an extension of pre-existing hypnozoite `density' models \parencite{white2014modelling, anwar2021multiscale} to accommodate a rigorous characterisation of superinfection. Under the deterministic model derived in \textcite{mehra2022superinf}, we can recover population-level distributions for various quantities of epidemiological interest without encountering the computational overheads that have curtailed previous efforts to model explicit hypnozoite densities \parencite{white2018mathematical}. The conceptual underpinning of the model detailed in \textcite{mehra2022superinf} is the within-host framework we introduced in \textcite{mehra2022hypnozoite}, which captures hypnozoite and superinfection dynamics as a function of mosquito-to-human transmission intensity, or the force of reinfection (FORI), in a general transmission setting. In the present paper, we adopt an analogous mathematical construction to \textcite{mehra2022superinf}, extending our previous work to allow for long-latency (temperate) strains and the acquisition of transmission-blocking and antidisease immunity. \\

This paper is structured as follows. Section \ref{sec::within_host_human} focuses on the 
characterisation of within-host dynamics as a function of the FORI. We begin by extending the open network of queues introduced in \textcite{mehra2022hypnozoite}, which describes the joint dynamics of superinfection and the hypnozoite reservoir, to include a discretised immunity level (Section \ref{sec::queue_network}); as observed in \textcite{mehra2022hypnozoite}, this immunity level is governed by a shot noise process, akin to a previous model of antibody dynamics we introduced in \textcite{mehra2021antibody}. Rather than solving for the state probabilities based on the Kolmogorov forward differential equations for the queueing network (Section \ref{sec::chapman_Kolmogorov}), we derive a time-dependent PGF as in our previous work \parencite{mehra2021antibody, mehra2022hypnozoite} (Section \ref{sec::queue_pgf}). Specific models for antidisease and transmission-blocking  immunity are proposed in Sections \ref{sec::antidisease_imm} and \ref{sec::tb_imm} respectively. Section \ref{sec::hybrid_models} concerns the construction of a hybrid transmission model \parencite{nasell2013hybrid, henry2020hybrid}, predicated on the coupling of expected host and vector dynamics. To recover the expected dynamics of the vector population, we consider the Kolmogorov forward differential equations for an underlying birth-death process (Section \ref{sec::vector_dynamics}); while observing that the within-host probability mass function (PMF) can be regarded as the expected population-level frequency distribution of hypnozoite, superinfection and immunity states \parencite{henry2020hybrid}. We then derive a infinite compartmental model to couple host and vector dynamics (Section \ref{sec::countable_ode}). Steady state analysis --- including the identification of a bifurcation parameter governing the existence of endemic equilibria (Section \ref{sec::steady_state}) and a sensitivity analysis of endemic equilibrium solutions (Section \ref{sec::sensitivity_analysis}) ----- is performed using the within-host distributions derived in \textcite{mehra2022hypnozoite}. To characterise transient population-level dynamics, we adopt the approach detailed in \textcite{mehra2022superinf} to derive a reduced system of IDEs --- comprising an integral equation for the immune-modulated probability of human-to-mosquito transmission (per bloodmeal), and a set of ordinary differential equations (ODEs) governing the number of (un)infected and latent mosquitoes over time (Section \ref{sec::hybrid_ide}). As a function of the FORI derived under the reduced system of IDEs, we recover population-level distributions for various quantities of epidemiological interest --- including the size of the (non)-latent hypnozoite reservoir; superinfection; the prevalence of clinical infection and the relative contribution of relapses to the infection burden --- using the distributions derived in \textcite{mehra2022hypnozoite}. Concluding remarks are made in Section \ref{sec::discussion}.

\section{Within-host human dynamics: hypnozoite accrual, superinfection and immunity} \label{sec::within_host_human}

We have previously derived the functional dependence between the FORI, and the joint dynamics of blood-stage infection and the hypnozoite reservoir by constructing an open network of infinite server queues \parencite{mehra2022hypnozoite}. Here, we extend the model detailed in \textcite{mehra2022hypnozoite} to allow for the acquisition of immunity. Following the approach detailed in Appendix C.3 of \textcite{mehra2022hypnozoite}, we assume that the within-host acquisition of immunity is described by a generalised shot noise process such that:
\begin{itemize}
    \item the clearance of each primary infection/relapse elicits a boost of unit magnitude;
    \item the lifetime of each boost is exponentially-distributed with mean $1/w$; and
    \item the overall immunity level is given by the cumulative sum of boosts over the entirety of an individual's infection history.
\end{itemize}
As noted in \textcite{mehra2022hypnozoite}, this discretised model of immunity can be considered a variation of the antibody model proposed in \textcite{mehra2021antibody}, in which the clearance of each primary infection/relapse elicits a boost of random  magnitude that is then subject to exponential decay at a fixed (deterministic) rate.\\

In Section \ref{sec::queue_network} below, we propose an open network of infinite server queues to capture the within-host dynamics of superinfection, hypnozoite accrual and immune acquisition \parencite{mehra2022hypnozoite}. The Kolmogorov forward differential equations governing the time evolution of the joint PMF for the network are stated in Section \ref{sec::chapman_Kolmogorov}. Instead of directly solving the the Kolmogorov forward differential equations (which comprise an infinite-dimensional set of ODEs), we derive a joint PGF for the state of the network following a similar approach to \textcite{mehra2022hypnozoite} (Section \ref{sec::queue_pgf}). Specific models for transmission-blocking and antidisease immunity are detailed in Sections \ref{sec::antidisease_imm} and \ref{sec::tb_imm} respectively. To elucidate the dynamics captured by our within-host model, we discuss an illustrative sample path in Section \ref{appendix::queue_sample_path}.

\subsection{An open network of infinite server queues} \label{sec::queue_network}

\begin{figure}
    \centering
    \includegraphics[width=\textwidth]{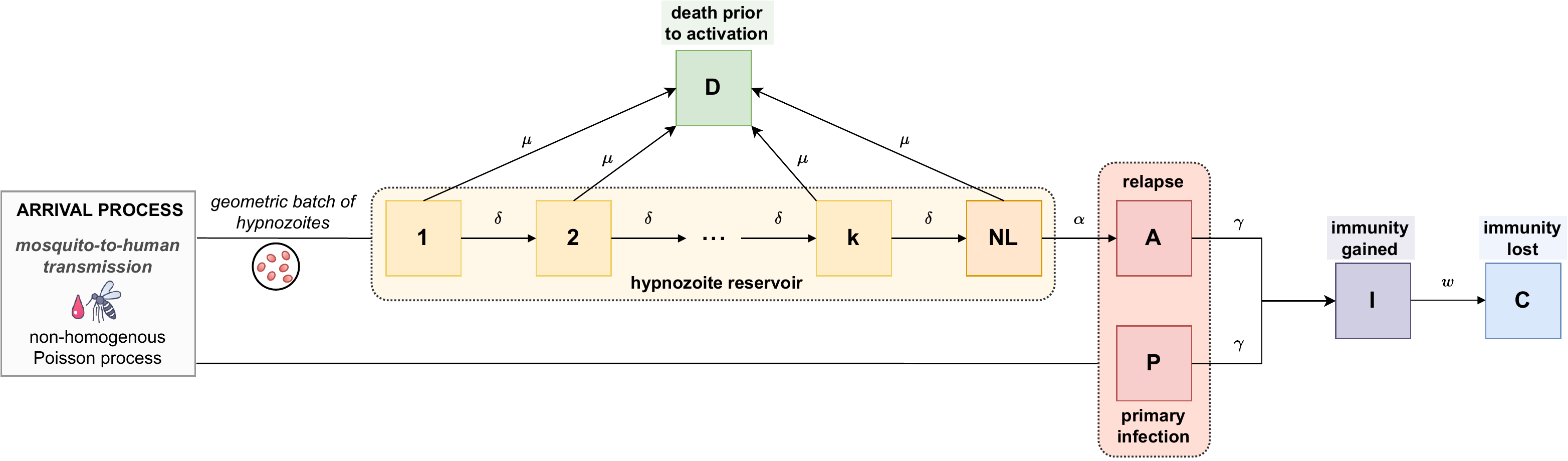}
    \caption{Schematic of open network of infinite server queues governing within-host hypnozoite and infection dynamics, allowing for the acquisition of immunity. Extended from \textcite{mehra2022hypnozoite} to include states related to immunity.}
    \label{fig:vivax_queue_imm}
\end{figure}

To capture the within-host acquisition of immunity, we extend the open network of queues detailed in \textcite{mehra2022hypnozoite} to include an additional node $I$, such that the occupancy of queue $I$ represents the immunity level. Specifically, we construct an open network of infinite server queues, labelled $1, \dots, k, NL, A, D, I, C, P$ (Figure \ref{fig:vivax_queue_imm}), where we define the compartments/nodes
\begin{itemize}
    \item $i \in \{1, \dots, k \}$ to represent hypnozoites that are present in latency compartment $i$ (that is, part of the hidden liver-stage reservoir but unable to activate)
    \item $NL$ to represent non-latent hypnozoites (that is, part of the hidden liver-stage reservoir and able to activate)
    \item $A$ to represent ongoing relapses from activated hypnozoites
    \item $D$ to represent hypnozoites that have died prior to activation
    \item $P$ to represent ongoing primary infections
    \item $I$ to represent cleared blood-stage infections (primary infections or relapses) that have given rise to an immunity increment of unit magnitude
    \item $C$ to represent the loss of immune memory.
\end{itemize}

As such, the state space for each hypnozoite is $S_h = \{ 1, \dots, k, NL, A, D, I, C\}$, while the state space for each primary infection is $S_p = \{ P, I, C \}$.\\

Arrivals into the network, which represent infective bites, are modelled to follow a non-homogeneous Poisson process with rate $\lambda(t)$. The consequences of each infective bite are two-fold:
\begin{itemize}
    \item a primary infection is immediately triggered, that is, a single ``customer'' enters queue $P$;
    \item a geometrically-distributed batch of hypnozoites (with mean size $\nu$) is established in the liver, entering latency compartment $1$ in the case of long-latency strains ($k>0$); and the non-latent compartment $NL$ in the case of short-latency strains ($k=0$).
\end{itemize}
Each hypnozoite/infection is assumed to flow independently through the network. Latent hypnozoites in the liver (that is, states $i \in \{1, \dots, k \}$) may either die at rate $\mu$, or shift to successive latency compartments at  rate $\delta$. This equates to exponentially-distributed service times, with mean duration $1/(\delta + \mu)$, in each of queues $i \in \{1, \dots, k \}$. A departure from queue $i$ is routed to either queue $D$ (representing hypnozoite death) with probability $\mu/(\delta  + \mu)$; or queue $(i+1)$ (representing progression to the next dormancy compartment) with probability $\delta/(\delta + \mu)$.\\

In contrast, non-latent hypnozoites (state $NL$) undergo death at rate $\mu$, and activation at rate $\alpha$.  As such, service times in queue $NL$ are modelled to be exponentially-distributed with mean $1/(\alpha + \mu)$, with departures routed either into queue $A$ (in which case hypnozoite activation has triggered a relapse) with probability $\alpha/(\alpha + \mu)$, or queue $D$ (in which case the hypnozoite has died prior to activation) with probability $\mu/(\alpha + \mu)$.\\
 
The clearance of each blood-stage infection is assumed to be independent, and modelled to occur at some constant rate $\gamma$, amounting to exponentially-distributed service times (with mean duration $1/\gamma$) in both queues $A$ and $P$ (representing relapses and primary infections respectively).\\

To capture the boosting of immunity with exposure, we assume that queue $I$ receives all departures from queues $A$ and $P$ (corresponding to cleared blood-stage infections); that is, an immune boost of unit magnitude is acquired upon the clearance of each primary infection or relapse. To capture the waning of immune memory with time, we assume that each immune boost is retained for an exponentially-distributed period of time with mean $1/w$ --- coinciding precisely with the service time in queue $I$. The number of busy servers in queue $I$ therefore acts as a measure of within-host immunity. All departures from queue $I$ are routed to queue $C$, where they remain indefinitely.\\

In a natural generalisation of this queueing network, the the stratification of blood-stage infection and immunity into different compartments could allow us to capture additional stages of the parasite lifecycle and further biological realism.

\subsubsection{Modelling correlates of immunity}

We can formulate correlates of immune protection as time-dependent functionals of the state of the open network, with the the host immunity level $N_I(t)$ mapped to the degree of immune protection at time $t$. To preserve the independence structure of the queueing network, however, these functionals cannot have any direct feedback into the within-host model. This limits the forms of immunity that are analytically tractable under our model.\\

A key assumption of the within-host model is that the arrival process is independent of the state of the network. As such, we cannot capture pre-erythrocytic immunity, which modulates the probability of successful mosquito-to-human transmission, whereby the arrival process, comprising mosquito bites, would depend on the number of busy servers $N_I(t)$ in node $I$. The assumption of independence between service rates within each node and the state of queueing network is equally important. We are thus unable to account for the potential modulation of (blood-stage) parasite clearance rates --- or equivalently, the service rate for nodes $A$ and $P$ --- as a function of the host immunity level $N_I(t)$, that is, the state of node $I$. We could, however, introduce \textit{deterministic} time variation in the rate of clearance of blood-stage infection $\gamma$ to model age-related physiological factors.\\

Immune correlates that are amenable under our within-host framework include:
\begin{itemize}
    \item The probability of exhibiting clinical symptoms or high-density parasitemia, as a manifestation of antidisease immunity (Section \ref{sec::antidisease_imm}).
    \item The probability of human-to-mosquito transmission when an uninfected mosquito takes a bloodmeal from a blood-stage infected human host, as a measure of transmission-blocking immunity (Section \ref{sec::tb_imm}).
\end{itemize}

There is evidence to suggest that these forms of immunity are acquired on different time scales. Transmission-blocking immune memory, for instance, is believed to be relatively short-lived, with boosting driven largely by successive blood-stage infections in intervals of $<4$ months \parencite{gamage1992transmission}; antidisease immunity, in contrast, is believed to be more robust and longer-lived \parencite{mueller2013natural}. By augmenting the rate of decay of the probability of symptomatic blood-stage infection (antidisease) as a function of $N_I(t)$, relative to the probability of human-to-mosquito transmission (transmission-blocking), we allow for strong antidisease protection to develop more rapidly than transmission-blocking protection, and be maintained at lower transmission intensities. Before discussing these immune correlates, however, we derive an analytic expression for the distribution of the state of the queueing network at time $t$ \parencite{mehra2022hypnozoite}.

\subsection{Kolmogorov forward differential equations} \label{sec::chapman_Kolmogorov}

Denote by $N_s(t)$ the number of hypnozoites/infections in each state $s \in S:= S_h \cup S_p$ at time $t$ and set
\begin{align*}
    H_{i_1, \dots, i_k, i_{NL}, j, k}(t) = P(N_1(t) = i_1, \dots, N_k(t) = i_k, N_{NL}(t) = i_{NL}, N_A(t) + N_P(t) = j, N_I(t) = k).
\end{align*}

Then by the Kolmogorov forward differential equations, the time evolution of the state probabilities $H_{i_1, \dots, i_k, i_{NL}, j, k}(t)$ is governed by the countable system of ODEs\par\nobreak
\vspace{-4.5mm}
{\footnotesize\begin{align}
    &\frac{d H_{i_1, \dots , i_k, i_{NL},j,k}}{dt} = \underbrace{\lambda(t) \bigg[ -H_{i_1, \dots , i_k, i_{NL},j,k}(t) +  \sum^{i_1}_{\ell=0} \frac{1}{\nu + 1} \Big( \frac{\nu}{ \nu + 1} \Big)^{i_1 - \ell} H_{\ell, \dots , i_k, i_{NL},j-1,k}(t) \bigg]}_{\text{reinfection (geometric batch of hypnozoites + primary infection triggered)}} \notag \\
    & + \underbrace{\mu \bigg[ - \Bigg( \sum^k_{\ell=1} i_\ell + i_{NL} \Bigg) H_{i_1, \dots , i_k, i_{NL},j,k}(t) + \sum^k_{\ell=1} (i_\ell+1) H_{i_1, \dots , i_{\ell - 1}, i_\ell + 1, i_{\ell+1}, \dots i_k, i_{NL},j,k}(t) + (i_{NL}+1) H_{i_1, \dots , i_k, i_{NL} + 1,j,k}(t) \bigg]}_{\text{death of a hypnozoite in the liver (latent or non-latent) prior to activation}} \notag \\
    & + \underbrace{\delta \bigg[ - \sum^k_{\ell=1} i_\ell H_{i_1, \dots , i_k, i_{NL},j,k}(t) + \sum^{k-1}_{\ell=1} (i_\ell + 1) H_{i_1, \dots , i_{\ell - 1}, i_\ell + 1, i_{\ell+1} -1, \dots i_k, i_{NL},j,k}(t) + (i_{k}+1) H_{i_1, \dots i_k + 1, i_{NL} - 1,j,k}(t) \bigg]}_{\text{progression of a latent hypnozoite to the next latency compartment}} \notag \\
    &+ \underbrace{\alpha \Big[ - i_{NL} H_{i_1, \dots , i_k, i_{NL},j,k}(t) + (i_{NL}+1) H_{i_1, \dots , i_k, i_{NL}+1,j-1,k}(t) \Big]}_{\text{activation of a non-latent hypnozoite, triggering a relapse}} \notag \\
    &+ \underbrace{\gamma \Big[ -j H_{i_1, \dots , i_k, i_{NL},j,k}(t) +  (j+1) H_{i_1, \dots , i_k, i_{NL},j+1,k-1}(t) \Big]}_{\text{clearance of blood-stage infection + gain of immunity increment}} \notag \\
    &+ \underbrace{w \Big[ - k H_{i_1, \dots , i_k, i_{NL},j,k}(t) + (k+1) H_{i_1, \dots , i_k, i_{NL},j,k+1}(t) \Big]}_{\text{waning of immune memory}}. \label{human_Kolmogorov}
\end{align}}

Consider a human population of fixed size $P_H$, with each individual taken to be immune- and infection-naive at time zero. In the absence of demography (that is, birth/death), we can re-interpret the within-host PMF $H_{i_1, \dots, i_k, i_{NL}, j, k}(t)$ as the \textit{expected} proportion of humans with $i_m$ hypnozoites in state $m \in \{1, \dots, k, NL\}$; a blood-stage infection comprising $j$ parasite broods and immunity level $k$. Equation (\ref{human_Kolmogorov}) can therefore be thought to govern the \textit{expected} proportion of humans in each hypnozoite/infection/immunity state \parencite{henry2020hybrid}. In Section \ref{sec::countable_ode}, we will draw on Equation (\ref{human_Kolmogorov}) to construct a hybrid transmission model, comprising a countably infinite system of ODEs. Our aim in the present section, however, is to characterise the within-host PMF. While the infinite-dimensional system of ODEs given by Equation (\ref{human_Kolmogorov}) is difficult to solve, we can readily derive the joint PGF governing the state of the network following similar reasoning to \textcite{mehra2022hypnozoite}.

\subsection{Joint PGF for the state of the queue} \label{sec::queue_pgf}
Rather than solving Equation (\ref{human_Kolmogorov}) to yield the probability mass function (PMF) for the state of the queue directly, we derive a joint PGF for the state of the network from first principles, using an argument which is an extension of that in \textcite{mehra2022hypnozoite}:
\begin{align*}
     \EX \Big[ \prod_{s \in S} z_s^{N_s(t)} \Big] =  \sum^\infty_{i_1=0} \dots \sum^\infty_{i_k=0} \sum^\infty_{i_{NL}=0} \sum^\infty_{j=0} \sum^\infty_{k=0} z_1^{i_1} \cdot \dots \cdot z_k^{i_k} \cdot z_{NL}^{i_{NL}} \cdot z_A^{j} z_I^{k} \cdot H_{i_1, \dots, i_k, i_{NL}, j, k}(t).
\end{align*}
The PGF can be viewed as a transformation of the PMF. Since there is a one-to-one correspondence between PGFs and PMFs, the derivation of a joint PGF is sufficient to uniquely characterise the distribution of the queueing network. Under the assumption of geometrically-distributed batch arrivals, we can invert the marginal PGF to recover PMFs for quantities of epidemiological interest, as discussed in \textcite{mehra2022hypnozoite}.\\

Treating the dynamics of each hypnozoite/infection to be independent \parencite{harrison1981note}, we begin by characterising the probability mass function for a single hypnozoite/primary infection that enters the network at time zero. Here, we extend the activation-clearance model proposed by \textcite{white2014modelling} --- and discussed in detail in \textcite{mehra2020activation} --- to allow for the clearance of blood-stage infection (which was also examined in \textcite{mehra2022hypnozoite}) and the gain/loss of immunity (as introduced in the present manuscript). To characterise hypnozoite dynamics, we consider the flow of an arrival into either queue $NL$ (for short-latency strains) or queue $1$ (for long-latency strains) through the queueing network shown in Figure \ref{fig:vivax_queue_imm}. Similarly, the dynamics of each primary infection are described by the flow of an arrival into queue $P$ through the network.\\

Denote by $p_{h,s}(t)$ the probability that a hypnozoite established at time zero is in state $s \in S_h$ at time $t$. By the Kolmogorov forward differential equations, it follows that
\begin{align}
    \frac{dp_{h, 1}}{dt} &= -(\delta + \mu) p_{h,1}(t) \label{hyp1_Kolmogorov}\\
    \frac{dp_{h, \ell}}{dt} &= -(\delta + \mu) p_{h,\ell}(t) + \delta p_{h, \ell - 1}(t) \text{ for } \ell \in \{2, \dots, k \}  \\
    \frac{dp_{h, NL}}{dt} &= -(\alpha + \mu) p_{h,NL}(t) + \delta p_{h,k}(t) \\
    \frac{dp_{h, A}}{dt} &= -\gamma p_{h, A}(t) +  \alpha p_{h,NL}(t)  \\
    \frac{dp_{h, I}}{dt} &= -w p_{h,I}(t) + \gamma p_{h,A}(t) \label{hypI_Kolmogorov}\\
    \frac{dp_{h, C}}{dt} &= w p_{h,I}(t) \label{hypC_Kolmogorov}\\
    \frac{dp_{h, D}}{dt} &= \mu \Big( \sum^k_{i=1} p_{h,i}(t) + p_{h, NL}(t) \Big) \label{hypD_Kolmogorov}
\end{align}
with the initial condition
\begin{align}
    \mathbf{p}_h(0) = \begin{cases}
    (p_{h,1}(0), \dots, p_{h,k}(0), p_{h,NL}(0), p_{h,A}(0), p_{h,C}(0), p_{h,D}(0)) = (1, 0, \dots, 0, 0, 0, 0, 0) & \text{ if } k>0\\
    (p_{h, NL}(0), p_{h,A}(0), p_{h,C}(0), p_{h,D}(0)) = (1, 0, 0, 0) & \text{ if } k=0
    \end{cases}.\label{ic}
\end{align}

Following similar reasoning to \textcite{mehra2020activation, mehra2022hypnozoite}, we can solve the system of ODEs given by Equations (\ref{hyp1_Kolmogorov}) to (\ref{hypI_Kolmogorov}) analytically; solutions are given in Equations (\ref{l_eq}) to (\ref{i_eq}) in Appendix \ref{sec::single_hyp}. Since we are not interested in the the distribution of dead hypnozoites or cleared infections over time, we do not provide solutions to the ODEs (\ref{hypC_Kolmogorov}) and (\ref{hypD_Kolmogorov}).\\

Likewise, we can characterise the probabilistic time course for each primary infection. Denote by $p_{p,s}(t)$ the probability that a primary infection triggered at time zero is in state $s \in S_p$ at time $t$. We can solve the Kolmogorov forward equations
\begin{align*}
    \frac{dp_{p,P}}{dt} = -\gamma p_{p,P}(t) \qquad \qquad \frac{dp_{p,I}}{dt} = -w p_{p,I}(t) + \gamma p_{p,P}(t) \qquad \qquad
    \frac{dp_{p,C}}{dt} &= w p_{p,I}(t)
\end{align*}
with initial condition
\begin{align*}
    \mathbf{p}_p(0) = (p_{p,P}(0), p_{p,I}(0), p_{p,C}(0)) = (1, 0,0)
\end{align*}
to yield
\begin{align}
    p_{p, P} = e^{-\gamma t} \qquad \qquad p_{p, I} = \frac{\gamma}{\gamma - w} \Big( e^{-w t} - e^{-\gamma t} \Big) \qquad \qquad p_{p, C}(t) = 1 - p_{p,P}(t) - p_{p, I}(t). \label{p_eq}
\end{align}
Embedding these state probabilities in an epidemiological framework, as elucidated in \textcite{mehra2022hypnozoite}, the joint PGF for 
\begin{align*}
    \mathbf{N}(t) = (N_1(t), \dots, N_k(t), N_{NL}(t), N_A(t), N_D(t), N_I(t), N_C(t), N_P(t))
\end{align*}
is given by
\begin{align}
     G(t, & z_1, \dots, z_k, z_{NL}, z_A, z_D, z_C, z_I, z_{P}) := \EX \Big[ \prod_{s \in S} z_s^{N_s(t)} \Big] \notag\\
     &= \exp \bigg\{ - \int^t_0 \lambda(\tau) \bigg[ 1 -  \frac{\sum_{s \in S_p} z_s \cdot p_{p, s}(t - \tau)} {1 + \nu \big( 1- \sum_{s \in S_h} z_s \cdot p_{h,s}(t-\tau) \big)} \bigg] d \tau \bigg\}. \label{vivax_multi_pgf_imm}
\end{align}

In \textcite{mehra2022hypnozoite}, we recovered analytic expressions for the distributions of several biologically-relevant quantities --- encompassing the size of the (non)-latent hypnozoite reservoir; the number of parasite broods co-circulating in the bloodstream; the relative contribution of relapses to the infection burden and the cumulative number of recurrences (that is, primary infections and relapses) experienced over time --- using the joint PGF given by Equation (\ref{vivax_multi_pgf_imm}). Formulae relevant to the present manuscript are recapitulated in Appendix \ref{appendix::steady_state_dist}.

\subsection{Antidisease immunity} \label{sec::antidisease_imm}

While the number of broods co-circulating in the bloodstream at time $t$ is given by the total number of busy servers in nodes $A$ and $P$, that is, $N_A(t) + N_P(t)$, a large proportion of \textit{P. vivax} infections in endemic settings are asymptomatic, with implications for treatment and elimination strategies \parencite{almeida2018high, tadesse2018relative, ferreira2022relative}. The relative burden of (a)symptomatic blood-stage infection, which is a function of antidisease immunity, is therefore of epidemiological interest.\\

Conditional on the presence of blood-stage infection, we assume that the probability of an individual exhibiting clinical symptoms decreases by a factor of $p_c$ for each increment of immunity they harbour. As such, the probability of an individual with state $\mathbf{N}(t)$ exhibiting clinical symptoms is given by
\begin{align*}
    p_\text{clin}(t) = p_{c}^{N_I(t)} \cdot \mathbbm{1}_{\{ N_A(t) + N_P(t) > 0 \}}.
\end{align*}

Accounting for stochasticity in within-host hypnozoite and infection dynamics, the probability of an individual exhibiting clinical symptoms at time $t$ can be written
\begin{align}
   p_\text{clin}&(t) := \sum^\infty_{i_1=0} \dots \sum^\infty_{i_k=0} \sum^\infty_{i_{NL}=0}  \sum^\infty_{j=1} \sum^\infty_{k=0} p_{c}^k H_{i_1, \dots, i_k, i_{NL}, j, k}(t) \notag \\
   = & \EX \Big[ p_{c}^{N_I(t)} \big| N_A(t) + N_P(t) > 0 \Big] \cdot P \big( N_A(t) + N_P(t) > 0 \big) \notag \\
   = & \EX \Big[ p_{c}^{N_I(t)} \Big] - \EX \Big[ p_{c}^{N_I(t)} \big| N_A(t) + N_P(t) = 0 \Big]  \cdot P \big( N_A(t) + N_P(t) = 0 \big)  \label{c_imm_prelim}
\end{align}
using the law of total expectation. Setting $z_i = 1$ for $i \in S \setminus I$ in Equation (\ref{vivax_multi_pgf_imm}) to recover the marginal PGF for $N_I(t)$, we can write
\begin{align}
    \EX \Big[ p_{c}^{N_I(t)} \Big] = G(t, z_1=1, \dots, z_k=1, z_{NL}=1, z_A=1, z_D=1, z_C=1, z_I=p_{c}, z_{P}=1). \label{c_imm_exp_1}
\end{align}

By \textcite{xekalaki1987method}, we can recover the unnormalised PGF for $N_I(t)$, conditional on the absence of blood-stage infection (that is, $N_A(t) + N_P(t) = 0$), by firstly setting $z_i = 1$ for $i \in S \setminus \{ A, P, I \}$ in Equation (\ref{vivax_multi_pgf_imm}) (to recover the joint PGF for $(N_A(t), N_P(t), N_I(t))$, and then setting $z_A = z_P = 0$ (to exclusively consider the case $N_A(t) + N_P(t) = 0$), yielding the expression
\begin{align}
    \EX \Big[ & p_{c}^{N_I(t)} \big| N_A(t) + N_P(t) = 0 \Big]  \cdot P \big( N_A(t) + N_P(t) = 0 \big) \notag \\
    & = G(t, z_1=1, \dots, z_k=1, z_{NL}=1, z_A=0, z_D=1, z_C=1, z_I=p_{c}, z_{P}=0). \label{c_imm_exp_2}
\end{align}

Substituting Equations (\ref{c_imm_exp_1}) and (\ref{c_imm_exp_2}) into Equation (\ref{c_imm_prelim}), we recover an expression for $p_\text{clin}(t)$ as a function of the FORI $\lambda(\tau)$ in the interval $\tau \in [0, t)$:
\begin{align}
   p_\text{clin}(t) &= \exp \bigg\{ - \int^t_0 \lambda(\tau) I_M(\tau) \bigg[ 1 -  \frac{1 - (1-p_c) p_{h, I}(t - \tau)} {1 + \nu (1-p_c) p_{p, I}(t-\tau)} \bigg] d \tau \bigg\} \notag \\
    &  \qquad - \exp \bigg\{ - \int^t_0 \lambda(\tau) \bigg[ 1 -  \frac{1 - (1-p_c) \cdot p_{p, I}(t-\tau) - p_{p, A}(t-\tau) } {1 + \nu (1-p_c) p_{h, I}(t-\tau) + \nu p_{h, A}(t-\tau)} d \tau \bigg\} \label{prob_clin}
\end{align}
where we have used the joint PGF given by Equation (\ref{vivax_multi_pgf_imm}). In a similar vein, we can introduce analogous models linking the probabilities of (sub)microscopic parasitemia and detectability (through light microscopy versus rapid diagnostic tests versus qPCR assays) to the immunity level $N_I(t)$.

\subsection{Transmission-blocking immunity} \label{sec::tb_imm}

Here, we propose a model for transmission-blocking immunity by introducing a functional dependence between the probability of successful human-to-mosquito transmission and the immune status of a blood-stage infected individual.\\

For an immune-naive, blood-stage infected individual, we set the probability of successful human-to-mosquito transmission to be $p_0$.  We further assume that the probability of successful human-to-mosquito transmission is reduced by a factor of $p_{tb} \in [0, 1]$ for each increment of immunity. Suppose a mosquito takes a bloodmeal from a human with state $\mathbf{N}(t)$ at time $t$.  Conditional on the state of a human host $\mathbf{N}(t)$, we thus define the probability of successful human-to-mosquito transmission $p_{h \to m}(t)$ to be
\begin{align*}
    p_{h \to m}(t) = p_0 \cdot p_{tb}^{N_I(t)} \cdot \mathbbm{1}_{\{ N_A(t) + N_P(t) > 0 \}}.
\end{align*}

Under our stochastic epidemiological framework, following similar reasoning to Section \ref{sec::antidisease_imm}, we can recover the probability of successful human-to-mosquito transmission when a mosquito bites an individual at time $t$ as a function of the FORI $\lambda(\tau)$ in the interval $\tau \in [0, t)$ 
\begin{align}
   p_{h \to m}&(t) := p_0 \sum^\infty_{i_1=0} \dots \sum^\infty_{i_k=0} \sum^\infty_{i_{NL}=0}  \sum^\infty_{j=1} \sum^\infty_{k=0} p_{tb}^k H_{i_1, \dots, i_k, i_{NL}, j, k} \notag \\
   = & p_0 \Big( \EX \Big[ p_{tb}^{N_I(t)} \Big] - \EX \Big[ p_{tb}^{N_I(t)} \big| N_A(t) + N_P(t) = 0 \Big]  \cdot P \big( N_A(t) + N_P(t) = 0 \big) \Big) \notag \\
   =& p_0 \Big[ G(t, z_1=1, \dots, z_k=1, z_{NL}=1, z_A=1, z_D=1, z_C=1, z_I=p_{c}, z_{P}=1) \notag \\
    & \qquad \quad -  G(t, z_1=1, \dots, z_k=1, z_{NL}=1, z_A=0, z_D=1, z_C=1, z_I=p_{c}, z_{P}=0) \Big]  \notag\\
    =& p_0 \Bigg[  \exp \bigg\{ - \int^t_0 \lambda(\tau) \bigg[ 1 -  \frac{1 - (1-p_{tb}) p_{h, I}(t - \tau)} {1 + \nu (1-p_{tb}) p_{p, I}(t-\tau)} \bigg] d \tau \bigg\} \notag \\
    &  \quad - \exp \bigg\{ - \int^t_0 \lambda(\tau) \bigg[ 1 -  \frac{1 - (1-p_{tb}) \cdot p_{p, I}(t-\tau) - p_{p, A}(t-\tau) } {1 + \nu (1-p_{tb}) p_{h, I}(t-\tau) + \nu p_{h, A}(t-\tau)} d \tau \bigg\} \Bigg]. \label{prob_h_to_m}
\end{align}

Note that Equation (\ref{prob_h_to_m}) accounts for both the acquisition of immunity and the probability of blood-stage infection.\\

The quantity $p_{h \to m}(t)$ is of particular importance since it underpins the coupling between host and vector dynamics; the time evolution of the expected number of infected mosquitoes (and consequently, the FORI) is dependent only on $p_{h \to m}(t)$ and several (known) transmission parameters (see Section \ref{sec::vector_dynamics}). Equation (\ref{prob_h_to_m}) will be of particular use in Section \ref{sec::hybrid_ide}, where we construct a reduced hybrid transmission model.

\subsection{Illustrative sample path} \label{appendix::queue_sample_path}

\begin{figure}
    \centering
    \includegraphics[width=0.75\textwidth]{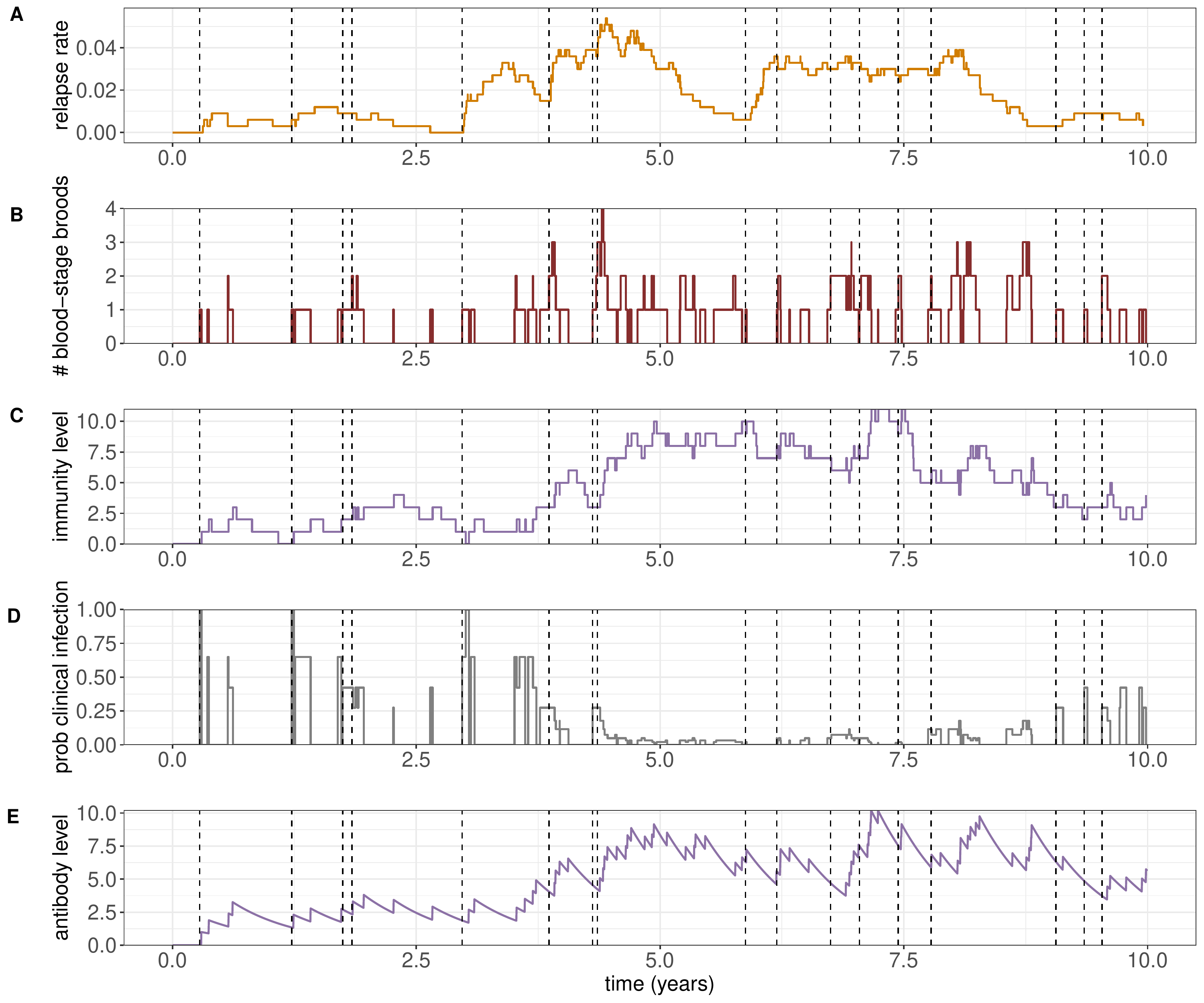}
    \captionsetup{singlelinecheck=off}
    \caption{Sample path, obtained through direct stochastic simulation, for an individual subject to a constant FORI ($\lambda = 2$ year$^{-1}$) over a period of 10 years. At time zero, the individual is immune-naive, and harbours neither liver- or blood-stage infection. Each infective bite (indicated with a dashed vertical line) triggers a primary infection, in addition to establishing a geometrically-distributed batch of hypnozoites in the liver, with mean size $\nu = 6.4$, as per \textcite{white2016variation}. We consider long-latency hypnozoites, with $k=2$ and $\delta = 1/100$ day$^{-1}$. Values for the hypnozoite activation $\alpha = 1/334$ day$^{-1}$ and death $\mu = 1/442$ day$^{-1}$ rates are drawn from \textcite{white2014modelling}. Blood-stage infections (primary and relapse) are assumed to be cleared at constant rate $\gamma = 1/24$ day$^{-1}$, as per estimates from \textcite{white2018plasmodium}. Under the discretised model of immunity, the lifetime of each immunity boost is assumed to be exponentially-distributed with mean duration $1/w = 250$ days, with the probability of clinical symptoms (conditional on the presence of blood-stage infection) assumed to decrease by a factor of $p_c=0.65$ for each increment of immunity. 
    \protect\begin{enumerate}
        \protect\item[(A)] The rate of relapse $\alpha N_{NL}(t)$.
        \protect\item[(B)] The number of parasite broods $N_A(t) + N_P(t)$ co-circulating in the bloodstream.
        \protect\item[(C)] The discretised immunity level $N_I(t)$.
        \protect\item[(D)] The probability of clinical infection $p_c^{N_I(t)} \mathbbm{1}_{\{ N_A(t) + N_P(t) > 0 \}}$.
        \protect\item[(E)] The antibody level, as per the model introduced in \textcite{mehra2021antibody}, whereby the clearance of each blood-stage infection is associated with a unit boost of immunity that then decays exponentially (deterministically) with rate $1/250$ day$^{-1}$.
    \protect\end{enumerate}}
    \label{fig:sample_path_const_trans}
\end{figure}

A simulated sample path illustrating temporal variation in the relapse rate, superinfection status (that is, the number of co-circulating parasite broods in the bloodstream) and immunity level, is shown in Figure \ref{fig:sample_path_const_trans}, as an extension of previous simulations presented in \textcite{mehra2021antibody, mehra2022hypnozoite}. We assume that the individual is both immune- and infection-naive at time zero. Over the course of 10 years, they are subject to 17 infective mosquito bites (shown with vertical dashed lines). Temporal variation in the relapse rate (Figure \ref{fig:sample_path_const_trans}A), which is proportional to the size of the non-latent hypnozoite reservoir, arises from the interplay between hypnozoite replenishment (through mosquito inoculation) and clearance (through either activation or death). Blood-stage infections include both primary infections (triggered immediately upon mosquito inoculation) and relapses (triggered by hypnozoite activation), with temporally proximate reinfection and/or hypnozoite activation events yielding multiple blood-stage infections (Figure \ref{fig:sample_path_const_trans}B). The discretised immunity level (shown in Figure \ref{fig:sample_path_const_trans}C) likewise fluctuates, with the clearance of each blood-stage infection eliciting a unit boost that is retained for an exponentially-distributed period of time. The probability of clinical infection (which serves as a correlate of antidisease immunity) decays geometrically with the discretised immunity level (Figure \ref{fig:sample_path_const_trans}D). For comparison, in Figure \ref{fig:sample_path_const_trans}E, we illustrate the simplest case of our previous (continuous) model of antibody dynamics \parencite{mehra2021antibody}, whereby the clearance of each blood-stage infection is associated with a unit boost of immunity that then decays exponentially (deterministically). Both the discrete (Figure \ref{fig:sample_path_const_trans}C) and continuous (Figure \ref{fig:sample_path_const_trans}E) models of immunity yield qualitatively similar results in this case.

\section{Hybrid transmission models: coupling expected host and vector dynamics} \label{sec::hybrid_models}

We now construct hybrid transmission models \parencite{nasell2013hybrid, henry2020hybrid} to couple host and vector dynamics. We restrict our attention to a homogeneously mixing population of humans and mosquitoes. While the size of the human population $P_H$ is taken to be fixed (with no age structure), we allow for time-variation in the size of the mosquito population (e.g. due to climactic variation or the implementation of vector-based control). Vector dynamics are detailed Section \ref{sec::vector_dynamics}.\\

In \textcite{mehra2022superinf}, for a simpler model structure --- accounting only for superinfection and short-latency hypnozoite accrual --- we began by constructing a Markov population process (with countably many types) to couple host and vector dynamics. Using the work of \textcite{barbour2012law}, we then recovered a deterministic (infinite) compartmental model as the functional law of large numbers; that is, we showed that the sample paths of the Markov process converged to a deterministic sample path in the infinite population size limit, granted the number of mosquitoes per human was held fixed. We noted in \textcite{mehra2022superinf}, however, that an identical deterministic compartmental model could be recovered under a ``hybrid approximation'', whereby host and vector dynamics are coupled through expected values, as per the construction of \textcite{nasell2013hybrid, henry2020hybrid}. This hybrid construction is the focus of Section \ref{sec::countable_ode}; by regarding the within-host PMF as the expected population-level frequency distribution \parencite{henry2020hybrid}, we recover a compartmental model (comprising an infinite-dimensional system of ODEs) that can be viewed as natural extension of the Ross-Macdonald model to allow for superinfection, hypnozoite accrual and immune acquisition. We characterise endemic equilibria for this compartmental model by drawing on results derived at the within-host level (Section \ref{sec::steady_state}), before performing a sensitivity analysis (Section \ref{sec::sensitivity_analysis}).\\ 

To characterise the transient dynamics of the hybridised system, we adopt the strategy we introduced in \textcite{mehra2022superinf}. Specifically, we collapse the infinite-dimensional compartmental model into a reduced system of IDEs --- with a set of ODEs governing the time evolution of the number of (un)infected and latent mosquitoes over time; and an integral equation governing the (transmission-blocking) immunity-modulated probability of successful human-to-mosquito tranmsmission (Section \ref{sec::hybrid_ide}). Based on the time evolution of the number of infected mosquitoes under the reduced system of IDEs, we can recover population-level distributions for several quantities of epidemiological interest using our derived within-host distributions.

\subsection{Vector dynamics: birth, death and infectivity} \label{sec::vector_dynamics}

Here, we characterise the dynamics of the vector population. We assume mosquito-dynamics are described by a Markovian birth-death process, whereby:
\begin{itemize}
    \item Mosquito births follow a time-dependent rate $\omega(t)$ (reflecting, for instance, climactic variation)
    \item Mosquito lifetimes are exponentially-distributed with mean duration $1/g$;
    \item Each mosquito bites humans (within a population of fixed size $P_H$) at a potentially time-varying rate $\beta(t)$ (reflecting, for instance, the relaxation/intensification of vector-based control measures);
    \item Following successful human-to-mosquito transmission (due to a bloodmeal from a blood-stage infected human), a mosquito undergoes sporogony at rate $\eta$;
    \item After sporogony has occured, a mosquito remains infective to humans until death.
\end{itemize}

A mosquito that is undergoing sporogony following successful human-to-mosquito transmission is hereafter considered to be latent.\\

Denote by $I_M(t)$, $L_M(t)$, $U_M(t)$ the expected number of infected, latent and uninfected mosquitoes respectively at time $t$. By the Kolmogorov forward differential equations for the Markovian birth-death process governing the vector population (see Appendix \ref{appendix:expected_vector} for details), we obtain the system of coupled ODEs
\begin{align}
    \frac{dI_M}{dt} &= \eta L_M(t) - g I_M(t) \label{d_im}\\
    \frac{dL_M}{dt} &= \beta(t) p_{h \to m}(t) U_M(t) - (g + \eta) L_M(t) \label{d_lm}\\
    \frac{dU_M}{dt} &= \omega(t) \big[ I_M(t) + L_M(t) + U_M(t) \big] - \big[ \beta(t) p_{h \to m}(t) + g \big] U_M(t). \label{d_um}
\end{align}
governing the time evolution of $I_M(t)$, $L_M(t)$, $U_M(t)$ as a function of the probability of human-to-mosquito transmission $p_{h \to m}(t)$ per bloodmeal.

\subsection{A countable system of ODEs} \label{sec::countable_ode}

Under a hybrid approximation, we seek to couple expected host and vector dynamics \parencite{nasell2013hybrid, henry2020hybrid}. Here, we recall two key observations: 
\begin{itemize}
    \item As we noted in Section \ref{sec::chapman_Kolmogorov}, for a human population of fixed size $P_H$, the system of ODEs given by Equation (\ref{human_Kolmogorov}) governs the \textit{expected} proportion of humans in each hypnozoite/infection/immunity state as a function of the FORI \parencite{henry2020hybrid}.
    \item The time evolution of the \textit{expected} number of (un)infected mosquitoes in the population is governed by Equations (\ref{d_im}) and (\ref{d_um}) conditional on the probability of successful human-to-mosquito transmission per bloodmeal.
\end{itemize}

To construct a hybrid transmission model, as per the approach of \textcite{nasell2013hybrid}, it remains to characterise the FORI, and the probability of human-to-mosquito transmission.\\

We assume that mosquito-to-human transmission occurs with fixed probability $p_{m \to h}$ when a human is bitten by an infected mosquito. We do not allow the parameter $p_{m \to h}$ to be modulated by immunity, that is, we do not account for the acquisition of pre-erythrocytic immunity. As a function of the number of infected $I_M(t)$ mosquitoes over time, the FORI can therefore be written
\begin{align}
    \lambda(t) =  \frac{\beta(t) p_{m \to h} I_M(t)}{P_H}. \label{fori}
\end{align}

Likewise, as a function of the proportion of humans $H_{i_1, \dots, i_k, i_{NL}, j, k}(t)$ with $i_{\ell}$ hypnozoites in state $\ell \in \{1, \dots, k, NL \}$; $j$ co-circulating broods in the bloodstream; and immunity level $k$ at time $t$, the probability of successful human-to-mosquito transmission $p_{h \to m}(t)$ can be written
\begin{align*}
    p_{h \to m}(t) = p_0 \sum^\infty_{i_1=0} \dots \sum^\infty_{i_k=0}  \sum^\infty_{i_{NL}=0} \sum^\infty_{j=1} \sum^\infty_{k=0} p_{tb}^k  H_{i_1, \dots , i_k, i_{NL},j,k}(t)
\end{align*}
as per the model of transmission-blocking immunity detailed in Section \ref{sec::tb_imm}.\\

Then following a similar approach to \textcite{nasell2013hybrid}, we recover the countable system of ODEs\par
\begin{figure}
    \centering
    \includegraphics[width=\textwidth]{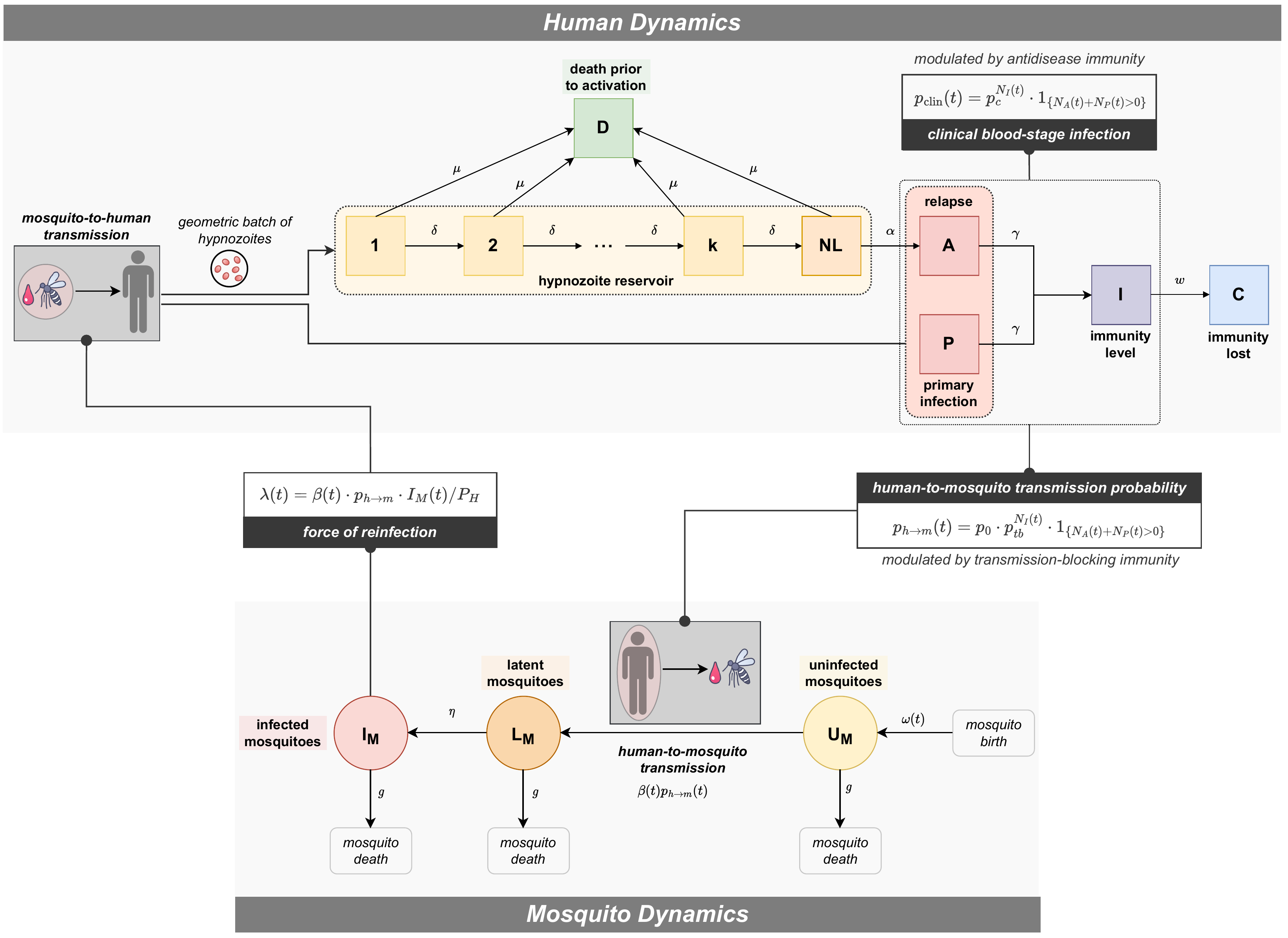}
    \caption{Schematic of hybrid transmission model structure, predicated on the coupling of expected host and vector dynamics \parencite{nasell2013hybrid, henry2020hybrid}. Here, the probabilistic distribution of the open network of infinite server queues governing within-host dynamics (Section \ref{sec::queue_network}) is re-interpreted as the expected proportion of humans in each hypnzoite/superinfection/immunity state. The coupling of host and vector dynamics is predicated on the force of reinfection $\lambda(t)$ (Equation (\ref{fori})), which is a function of the number of infected mosquitoes at time $t$; and the probability of successful human-to-mosquito transmission $p_{h \to m}(t)$ per bloodmeal (Equation (\ref{prob_h_to_m})), which is modulated both by the prevalence of blood-stage infection and the distribution of immunity in the human population at time $t$.}
    \label{fig:transmission_model_schematic}
\end{figure}

\nobreak
\vspace{-4.5mm}
{\footnotesize
\begin{align}
    &\frac{d H_{i_1, \dots , i_k, i_{NL},j,k}}{dt} = \beta(t) p_{m \to h} \frac{I_M(t)}{P_H}\bigg[ -H_{i_1, \dots , i_k, i_{NL},j,k}(t) +  \sum^{i_1}_{\ell=0} \frac{1}{\nu + 1} \Big( \frac{\nu}{ \nu + 1} \Big)^{i_1 - \ell} H_{\ell, \dots , i_k, i_{NL},j-1,k}(t) \bigg] \notag \\
    & + \mu \bigg[ - \sum^k_{\ell=1} i_\ell H_{i_1, \dots , i_k, i_{NL},j,k}(t) + \sum^k_{i=1} (i_\ell+1) H_{i_1, \dots , i_{\ell - 1}, i_\ell + 1, i_{\ell+1}, \dots i_k, i_{NL},j,k}(t) + (i_{NL}+1) H_{i_1, \dots , i_k, i_{NL} + 1,j,k}(t) \bigg] \notag \\
    & + \delta \bigg[ - \sum^k_{\ell=1} i_\ell H_{i_1, \dots , i_k, i_{NL},j,k}(t) + \sum^{k-1}_{i=1} (i_\ell + 1) H_{i_1, \dots , i_{\ell - 1}, i_\ell + 1, i_{\ell+1} -1, \dots i_k, i_{NL},j,k}(t) + (i_{k}+1) H_{i_1, \dots i_k + 1, i_{NL} - 1,j,k}(t) \bigg] \notag \\
    &+ \alpha \Big[ - i_{NL} H_{i_1, \dots , i_k, i_{NL},j,k}(t) + (i_{NL}+1) H_{i_1, \dots , i_k, i_{NL}+1,j-1,k}(t) \Big] \notag \\
    &+ \gamma \Big[ -j H_{i_1, \dots , i_k, i_{NL},j,k}(t) +  (j+1) H_{i_1, \dots , i_k, i_{NL},j+1,k-1}(t) \Big] \notag \\
    &+ w \Big[ - k H_{i_1, \dots , i_k, i_{NL},j,k}(t) + (k+1) H_{i_1, \dots , i_k, i_{NL},j,k+1}(t) \Big] \label{hybrid_human_ode} \\
    &\frac{dI_M}{dt} = \eta L_M(t) - gI_M(t) \label{hybrid_inf_mos_ode} \\
    &\frac{dL_M}{dt} = \beta(t) p_0 \sum^\infty_{i_1=0} \dots \sum^\infty_{i_k=0}  \sum^\infty_{i_{NL}=0} \sum^\infty_{j=1} \sum^\infty_{k=0} p_{tb}^k  H_{i_1, \dots , i_k, i_{NL},j,k}(t) U_M(t) - (g + \eta) L_M(t) \label{hybrid_latent_mos_ode} \\
    & \frac{dU_M}{dt} = \omega(t) ( I_M(t) + L_M(t) + U_M(t)) - \bigg( g + \beta(t) p_0 \sum^\infty_{i_1=0} \dots \sum^\infty_{i_k=0}  \sum^\infty_{i_{NL}=0} \sum^\infty_{j=1} \sum^\infty_{k=0} p_{tb}^k  H_{i_1, \dots , i_k, i_{NL},j,k}(t) \bigg) U_M(t)\label{hybrid_mos_ode}.
\end{align}}%
where we have used Equation (\ref{human_Kolmogorov}) and Equations (\ref{d_im}) to (\ref{fori}). A schematic of this model is provided in Figure \ref{fig:transmission_model_schematic}.\\

The compartmental model given by Equations (\ref{hybrid_human_ode}) and (\ref{hybrid_mos_ode}) represents a natural extension of the Ross-Macdonald framework to allow for hypnozoite accrual, superinfection and transmission-blocking immunity. We can also view Equations (\ref{hybrid_human_ode}) and (\ref{hybrid_mos_ode}) as an extension of the transmission model proposed by \textcite{white2014modelling} to allow for long-latency hypnozoites, immunity and explicit superinfection dynamics (as opposed to the approximation of superinfection through an appropriate recovery rate). A summary of model parameters, and their respective interpretations, is provided in Table \ref{table:model_params}.

\begin{footnotesize}
\begin{sidewaystable}
\centering
\begin{tabular}{|c|l|c|c|} 
 \hline
 \textbf{Parameter} & \textbf{Interpretation} & \textbf{Value} & \textbf{Source} \\
 \hline\hline
 $\alpha$ & Hypnozoite activation rate & $1/334$ day$^{-1}$ & \textcite{white2014modelling} \\ 
 \hline
 $\mu$ & Hypnozoite death rate & $1/442$ day$^{-1}$ & \textcite{white2014modelling} \\ 
 \hline
 $\delta$ & Rate of progression through hypnozoite latency compartments & $1/100$ day$^{-1}$ & assumed \\ 
 \hline
 $k$ & Number of hypnozoite latency compartments & $0,1,2$ & assumed  \\ 
 \hline
 $\nu$ & Mean number of hypnozoites established per bite & $6.4$ & \textcite{white2016variation} \\ 
 \hline
 $\gamma$ & Rate at which each blood-stage infection is cleared & $1/24$ day$^{-1}$ & \textcite{white2018plasmodium}  \\
 \hline
 $w$ & Rate at which each immune increment/boost is lost & $1/250$ day$^{-1}$ & assumed \\
 \hline
 $p_{c}$ & Factor by which the probability of clinical/symptomatic & various & assumed \\
 & blood-stage infection decreases per unit level of immunity & &  \\ 
 \hline
 $p_0$ & Probability of human-to-mosquito transmission when a mosquito & 0.25, 0.65 & assumed\\
 & bites a blood-stage infected, immune-naive human & & \\ 
 \hline
 $p_{tb}$ & Factor by which the probability of human-to-mosquito  & $0.9$ & assumed \\
 & transmission decreases per unit level of immunity when a  & & \\ 
 & mosquito bites a blood-stage infected human & & \\ 
 \hline
 $p_{h \to m}(t)$ & Probability of human-to-mosquito transmission when an  & calculated & Equation (\ref{prob_h_to_m}) \\
 & uninfected mosquito takes a bloodmeal at time $t$ & & \\ 
  \hline
 $p_{m \to h}$ & Probability of mosquito-to-human transmission when  & 0.25 & \textcite{white2018mathematical}\\
 & an infected mosquito bites a human & & \\ 
 \hline
 $\beta(t)$ & Bite rate per mosquito & 0.21 day$^{-1}$ & \textcite{garrett1964human} \\ 
 \hline
 $g$ & Mosquito death rate & $0.1$ day$^{-1}$ & \textcite{gething2011modelling} \\ 
 \hline
 $\omega(t)$ & Mosquito birth rate at time $t$ & various  & assumed \\ 
 \hline
 $1/\eta$ & Mean duration of sporogony & $12$ days  & \textcite{gething2011modelling} \\ 
 \hline
 $\frac{P_M}{P_H}$ & Ratio of mosquito and human population size assuming $\omega(t)=g$ & 1.2 & assumed\\
 \hline
 $\lambda(t)$ & Force of reinfection (FORI) at time $t$ & calculated & Equation (\ref{fori}) \\
 \hline
\end{tabular}
\caption{Summary of model parameters.}
\label{table:model_params}
\end{sidewaystable}
\end{footnotesize}

\subsubsection{The stationary solution} \label{sec::steady_state}
Here, we seek to characterise steady state solutions to the system of ODEs given by Equations (\ref{hybrid_human_ode}) to (\ref{hybrid_mos_ode}). As such, we restrict ourselves to a setting where:
\begin{itemize}
    \item The bite rate per mosquito $\beta(t) = \beta$ remains constant over time; and
    \item The mosquito population size $I_M(t) + L_M(t) + U_M(t)= P_M$ is fixed, that is, the birth rate $\omega(t) = g$ exactly balances the (constant) death rate.
\end{itemize}

Denote by $H^*_{i_1, \dots, i_k, i_{NL}, j,k}$, $U^*_M$, $L^*_M$ $I^*_M$ the stationary solution to the system of IDEs given by Equations (\ref{hybrid_human_ode}) to (\ref{hybrid_mos_ode}), recovered by setting all time derivatives to zero.\\

We focus on the quantities $H^*_{i_1, \dots, i_k, i_{NL}, j,k}$ and $I^*_M$ since the overarching effect of sporogony is to introduce a scaling factor $g/(g+\eta)$ in the fraction of mosquitoes that, in the event of successful mosquito-to-human transmission, survive latency to transition from an uninfected to infected state. Setting the time derivative in Equation (\ref{hybrid_inf_mos_ode}) to zero and using the assumption of a fixed mosquito population size, we can formulate the number of latent $L_M^*$ and uninfected $U_M^*$ at steady state as functions of $I_M^*$ and $P_M$ as follows:
\begin{align*}
    L_M^* = \frac{g}{\eta} I_M^* \qquad U_M^* = P_M - \Big( 1 + \frac{g}{\eta} \Big) I_M^*.
\end{align*}

We observe that the disease (and immunity) free equilibrium $H^*_{0, \dots, 0, 0, 0, 0}=1, I^*_M=0$ always exists, Here, we seek to characterise the existence of endemic equilibrium solutions.\\

We begin by setting the time derivatives in Equations (\ref{hybrid_inf_mos_ode}) and (\ref{hybrid_mos_ode}) to zero to yield
\begin{align}
    \sum^\infty_{i_1=0} \dots \sum^\infty_{i_k=0}  \sum^\infty_{i_{NL}=0} \sum^\infty_{j=1} \sum^\infty_{k=0} p_{tb}^k  H^*_{i_1, \dots , i_k, i_{NL},j,k} = \frac{g I_M^*}{\beta p_0 \big( \frac{P_M}{1+g/\eta} -I^*_M \big)}. \label{im_sim_eq_2}
\end{align}

We then note that Equation (\ref{hybrid_human_ode}) is precisely the set of Kolmogorov forward differential equations for the open network of infinite server queues introduced in Section \ref{sec::within_host_human}. The PGF for the stationary limiting distribution of this queueing network, given a constant FORI $\lambda(t) = \beta p_{m \to h} I^*_M/P_H$, can be recovered by taking the limit $t \to \infty$ in Equation (\ref{vivax_multi_pgf_imm}). Therefore, using Equation (\ref{prob_h_to_m}) --- which we derived from Equation (\ref{vivax_multi_pgf_imm}) in Section \ref{sec::tb_imm} --- we suggest that
\begin{align}
     \sum^\infty_{i_1=0} & \dots \sum^\infty_{i_k=0}  \sum^\infty_{i_{NL}=0} \sum^\infty_{j=1} \sum^\infty_{k=0} p_{tb}^k  H^*_{i_1, \dots , i_k, i_{NL},j,k} \notag\\
     =& p_0 \Bigg( \exp \bigg\{ - \frac{\beta p_{m \to h} I^*_M}{P_H} \int^\infty_0  \bigg[ 1 -  \frac{1 - (1-p_{tb}) p_{h, I}(\tau)} {1 + \nu (1-_{tb}) p_{p, I}(\tau)} \bigg] d \tau \bigg\} \notag \\
    &  \qquad - \exp \bigg\{ - \frac{\beta p_{m \to h} I^*_M}{P_H} \int^\infty_0 \bigg[ 1 -  \frac{1 - (1-p_{tb}) \cdot p_{p, I}(\tau) - p_{p, A}(\tau) } {1 + \nu (1-p_{tb}) p_{h, I}(\tau) + \nu p_{h, A}(\tau)} d \tau \bigg\} \Bigg). \label{im_sim_eq_1}
\end{align}

Using a simple geometric argument (Appendix \ref{appendix::endemic_eq}), we can show that Equations (\ref{im_sim_eq_2}) and (\ref{im_sim_eq_1}) have at most one non-zero intersection (corresponding to an endemic equilibrium solution), and that this intersection exists if and only if
\begin{align}
    R_0^2 := \frac{\beta^2 p_0 p_{m \to h} P_M}{g (1 + g/\eta) P_H} \int^\infty_0 \bigg[ \frac{1 - (1-p_{tb}) p_{h, I}(\tau)} {1 + \nu (1-p_{tb}) p_{p, I}(\tau)}  - \frac{1 - (1-p_{tb}) \cdot p_{p, I}(\tau) - p_{p, A}(\tau) } {1 + \nu (1-p_{tb}) p_{h, I}(\tau) + \nu p_{h, A}(\tau)} \bigg]  d \tau > 1. \label{eq:R0}
\end{align}

Assuming that $R_0 >1$ (Equation (\ref{eq:R0})), an endemic equilibrium solution necessarily exists. As a function of the FORI $\lambda^* =\beta p_{m \to h} I^*_M/P_H$ at the endemic equilibrium, which is a function of the non-trivial solution $I_M^* \in (0, P_M]$ to Equations (\ref{im_sim_eq_2}) and (\ref{im_sim_eq_1}), we can recover population-level distributions for various quantities of epidemiological interest using the stationary limiting PGF recovered by setting $\lambda(t) = \lambda^*$ for all $t \geq 0$ and taking the limit $t \to \infty$ in Equation (\ref{vivax_multi_pgf_imm}). Relevant formulae (based on the derivations presented in \textcite{mehra2022hypnozoite})) are provided in Appendix \ref{appendix::steady_state_dist}.

\subsubsection{Sensitivity analysis for endemic equilibrium solutions} \label{sec::sensitivity_analysis}

We now perform a sensitivity analysis for the endemic equilibrium solutions. In Section \ref{sec::sens_no_imm}, we examine endemic equilibrium solutions in the absence of immunity. Endemic equilibria, allowing for transmission-blocking and antidisease immunity, are detailed in Section \ref{sec::imm_short_latency}.

\paragraph{Short-latency vs long-latency strains in the absence of transmission-blocking immunity} \label{sec::sens_no_imm}

We begin by examining steady state solutions for both short-latency ($k=0$) and long-latency ($k>0$) strains in the absence of transmission-blocking immunity ($p_{tb}=1$).\\

\begin{figure}
    \centering
    \includegraphics[width=\textwidth]{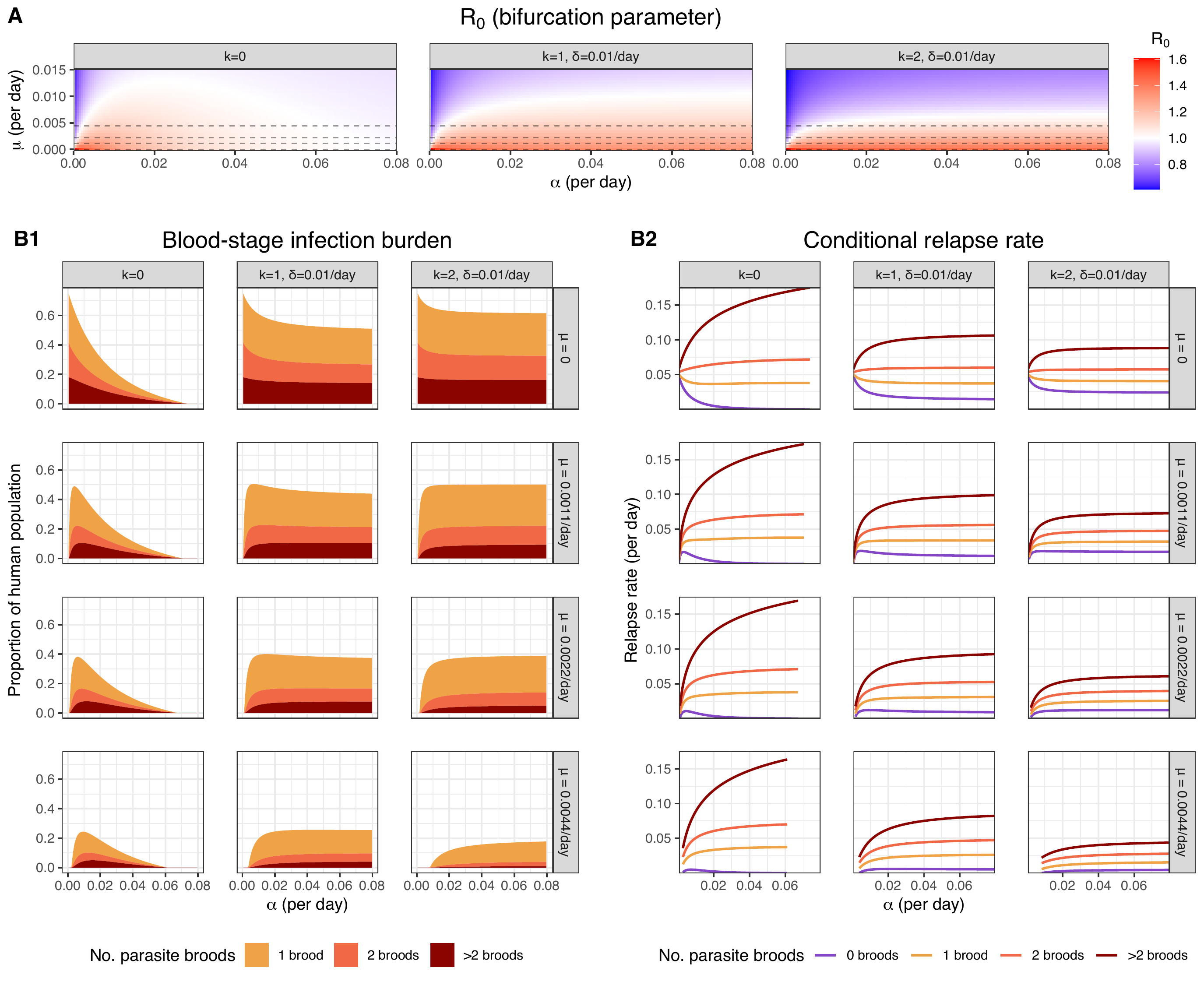}
    \captionsetup{singlelinecheck=off}
    \caption{Three-way sensitivity analysis (with respect to $\mu$, $\alpha$ and $k$) for steady state solutions solutions in the absence of transmission-blocking immunity ($p_{tb}=1$) for both short-latency ($k=0$) and long-latency ($k>0$) strains.
    {\protect\begin{enumerate}
        \protect\item[(A)] Sensitivity analysis for $R_0$ (Equation (\ref{eq:R0})). Here, we consider $\mu \in [0, 0.015)$ day$^{-1}$ and $\alpha \in [0, 0.08)$ day$^{-1}$, each in increments of $0.00055$ day$^{-1}$; and $k=0,1,2$, while fixing $\delta=1/100$ day$^{-1}$. Values of $\mu$ considered in (B) are indicated with dashed horizontal lines.
        \protect\item[(B)] Sensitivity analysis for (B1) the number of co-circulating parasite broods (Equations (\ref{moi_0}) to (\ref{moi_2})); and (B2) the relapse rate conditional on superinfection status (Equations (\ref{rel_rate_moi_0}) to (\ref{rel_rate_moi_3+})) at the endemic equilibrium solution. The prevalence of blood-stage infection and the number of infected mosquitoes $I_M^*$ at the endemic equilibrium are given by the non-trivial solution to Equations (\ref{im_sim_eq_1}) and (\ref{im_sim_eq_2}) (which exists, and is unique, iff $R_0>1$); endemic equilibrium solutions for quantities of epidemiological interest are recovered as functions of $I_M^*$. Here, we consider $\mu \in \{0, 0.0011, 0.0022, 0.0044\}$ day$^{-1}$; $\alpha \in [0, 0.08)$ day$^{-1}$ in increments of $0.00055$ day$^{-1}$; and $k=0,1,2$, while fixing $\delta=1/100$ day$^{-1}$.
    \protect\end{enumerate}}
    We set $P_M/P_H=1.2$, $p_0=0.25$ and parameters $\gamma, \nu, g = \omega(t), \beta, p_{m \to h}$ as per Table \protect\ref{table:model_params}.}
    \label{fig:no_imm_ss}
\end{figure}

Figure \ref{fig:no_imm_ss}A depicts a sensitivity analysis for the bifurcation parameter $R_0$ (Equation (\ref{eq:R0})) with respect to the hypnozoite activation $\alpha$, death $\mu$ and latency $k$ parameters. Recall that an endemic equilibrium exists, and is unique, if $R_0 > 1$; if $R_0<1$, only the disease-free equilibrium exists. The bifurcation boundary $R_0 = 1$ for parameters $(\alpha, \mu)$ is shown in white in Figure \ref{fig:no_imm_ss}A. In the absence of hypnozoite accrual ($\nu = 0$), no endemic equilibria exist for the set of transmission parameters considered here; the existence of endemic equilibrium solutions is therefore contingent on the relapse burden. The interplay between hypnozoite activation $\alpha$ and death $\mu$ rates governs the expected number of relapses per bite $\alpha/(\alpha + \mu)$ \parencite{white2014modelling}. As such, when the hypnozoite activation rate $\alpha$ is low relative to the hypnozoite death rate $\mu$, there are insufficient relapses to sustain transmission and no endemic equilibrium solution exists, that is, $R_0 < 1$ (Figure \ref{fig:no_imm_ss}A).\\

In the case of short-latency strains ($k=0$), we further observe that excessively high activation rates $\alpha$ preclude the existence of endemic equilibrium solutions (that is, yield $R_0 < 1$) (Figure \ref{fig:no_imm_ss}A); similar observations have been posited by \textcite{white2016variation} and \textcite{anwar2021multiscale}, albeit in the absence of superinfection. Without an enforced dormancy period, an elevated activation rate $\alpha$ gives rise to a high risk of relapse immediately after each infective bite. The rapid depletion of the hypnozoite reservoir following each bite --- driven by temporally proximate hypnozoite activation events --- leads to a divergence in relapse risk conditional on status of blood-stage (super)infection (Figure \ref{fig:no_imm_ss}B2). To justify why a high relapse rate for (blood-stage) superinfected individuals is a weaker driver of onward transmission than a high relapse rate for blood-stage uninfected individuals, we observe that the expected time to clearance for $j$ parasite broods is $\gamma(1 + 1/2 + \dots + 1/j)$; as such, an additional relapse for an individual with $m$ pre-existing broods in their bloodstream increases the expected time to (blood-stage) recovery by an increment of $\gamma/(m+1)$. We deduce that the stratification of relapse risk, conditional on the status of blood-stage infection, is driven by the time to the most recent infective bite in the case of fast-activating short-latency hypnozoites: while recently-inoculated individuals experience a high burden of both liver- and blood-stage infection, there is a limited burden of liver- and blood-stage infection \textit{between} successive mosquito bites, yielding a population-level reduction in the overall burden of blood-stage infection (Figure \ref{fig:no_imm_ss}B1). Hence, for elevated activation rates $\alpha$, there is a limited window of time following each infective bite for which an individual remains blood-stage infected, and therefore infective to mosquitoes; if the activation rate $\alpha$ is sufficiently high, then these windows of human-to-mosquito infectivity may be insufficient to sustain transmission in the steady state, in which case $R_0<1$ and no endemic equilibrium solution exists (Figure \ref{fig:no_imm_ss}A).\\

For long-latency strains ($k>0$), stochasticity in the enforced dormancy period prevents excessive overlap between hypnozoite activation events arising from the same bite, thereby reducing the sensitivity of the endemic equilibrium burden of blood-stage infection to elevated hypnozoite activation rates $\alpha$ (Figure \ref{fig:no_imm_ss}B1). Decreasing the variance of the dormancy period $k/\delta^2$, whilst fixing the expected duration $k/\delta$, would presumably increase the sensitivity of endemic equilibria to the hypnozoite activation rate $\alpha$, since hypnozoites would be more likely to emerge from dormancy at similar times. We observe that the assumption of independent dormancy, introduced in \textcite{mehra2020activation}, underpins this observation for long-latency strains; the collective dormancy assumption of \textcite{white2014modelling} --- whereby synchronicity in the latency phase means that hypnozoites established through the same infective bite emerge collectively from dormancy --- leads to greater sensitivity of endemic equilibrium solutions to elevated hypnozoite activation rates $\alpha$. Indeed, under a `binary' hypnozoite model predicated implicitly on the assumption of collective dormancy, \textcite{white2016variation} predict stronger constraints on the hypnozoite activation rate $\alpha$ than we predict in Figure \ref{fig:no_imm_ss} under the assumption of independent dormancy.\\

Elevated hypnozoite activation rates $\alpha$, however, give rise to a stratification of relapse risk by superinfection status, even in the case of long-latency strains ($k>0$) (Figure \ref{fig:no_imm_ss}B2). In the absence of hypnozoite death (that is, $\mu=0$), the burden of blood-stage infection at the endemic equilibrium is maximised for low hypnozoite activation rates $\alpha$ (Figure \ref{fig:no_imm_ss}B1), which yield broad temporal relapse distributions for each infective bite, and a population-level relapse risk that does not vary strongly by superinfection status (Figure \ref{fig:no_imm_ss}B2). For non-zero death rates $\mu>0$, however, hypnozoite death during the enforced dormancy period --- during which activation is prohibited --- serves as a key constraint. As such, elevated activation rates $\alpha$ (up to a point) yield an increasing burden of blood-stage infection for long-latency strains ($k>0$, Figure \ref{fig:no_imm_ss}B1), even as the risk of relapse stratified by superinfection status becomes more unbalanced and proportionately higher for individuals with pre-existing blood-stage infections (Figure \ref{fig:no_imm_ss}B2).

\paragraph{Short-latency strains with transmission-blocking immunity} \label{sec::imm_short_latency}

\begin{figure}
    \centering
    \includegraphics[width=0.8\textwidth]{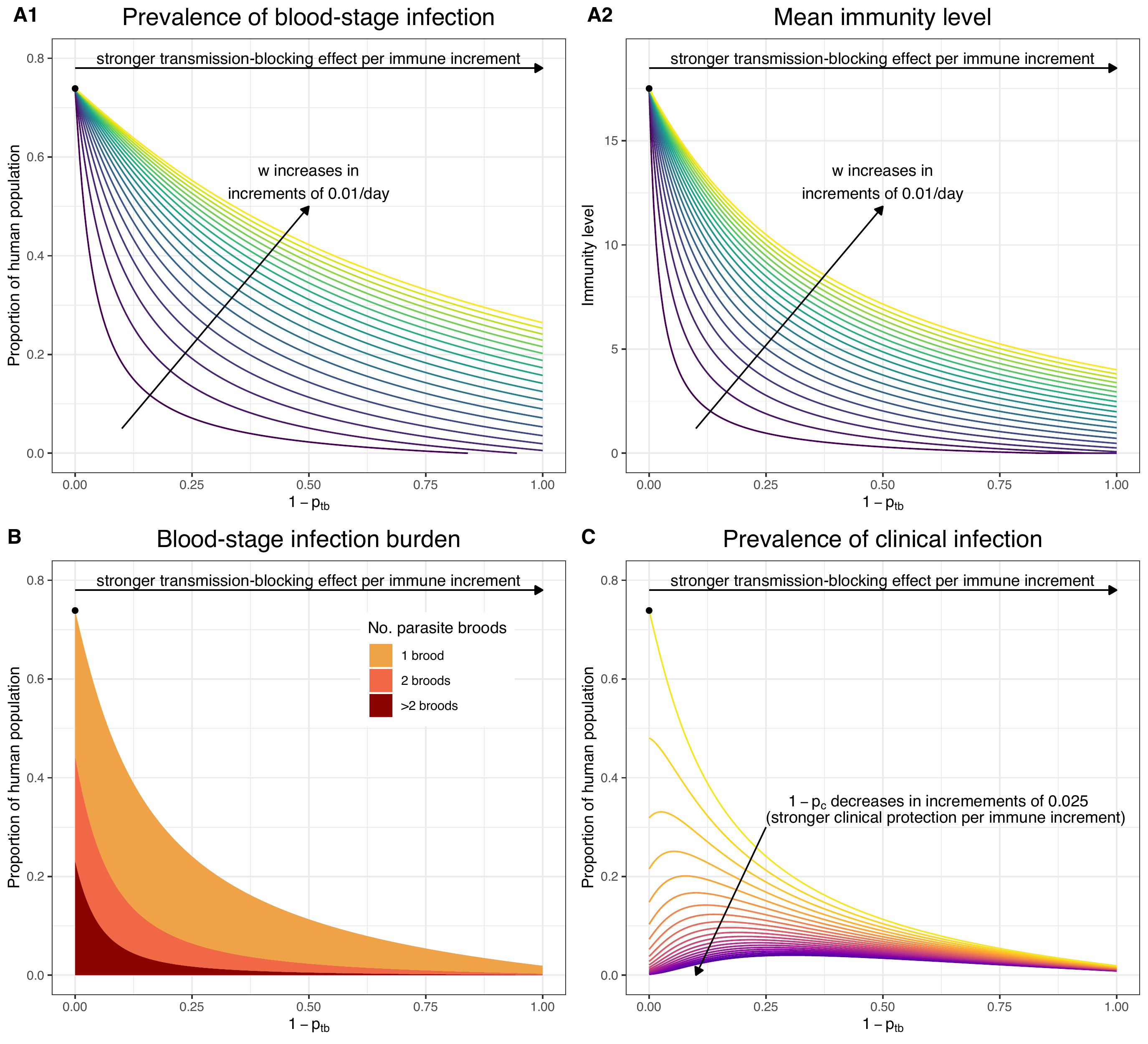}
    \captionsetup{singlelinecheck=off}
    \caption{Endemic equilibrium solutions for for short-latency strains ($k=0$) allowing for transmission-blocking immunity ($0 < (1 - p_{tb}) \leq 1$). The probability of human-to-mosquito transmission (per bloodmeal) and the number of infected mosquitoes $I_M^*$ at the endemic equilibrium are given by the non-trivial solution to Equations (\ref{im_sim_eq_1}) and (\ref{im_sim_eq_2}) (which exists, and is unique, iff $R_0>1$). Endemic equilibrium solutions for quantities of epidemiological interest are recovered as functions of $I_M^*$.
    \protect\begin{enumerate}
        \protect\item[(A)] Two-way sensitivity analysis (with respect to $(1-p_{tb})$ and $w$) for (A1) prevalence of blood-stage infection (Equation (\ref{moi_0})) and (A2) the mean immunity level (Equation (\ref{exp_imm})) at the endemic equilibrium. Here, we consider $w \in [0.001, 0.02)$ day$^{-1}$ in increments of $0.001$ day$^{-1}$ and $(1-p_{tb}) \in [0, 1)$ in increments of $0.005$.
        \protect\item[(B)] One-way sensitivity analysis (with respect to $p_{tb}$) for the number of co-circulating parasite broods in the bloodstream (Equations (\ref{moi_0}) to (\ref{moi_2})), with $w=1/250$ day$^{-1}$ fixed and $(1-p_{tb}) \in [0,1)$ in increments of $0.005$.
        \protect\item[(C)] Two-way sensitivity analysis (with respect to $p_{tb}$ and $p_c$) for the prevalence of clinical infection (Equation (\ref{prob_clin})) at the endemic equilibrium, where we consider $(1-p_{tb}) \in [0, 1)$ in increments of $0.005$ and $p_c \in [0.5, 1)$ in increments of $0.025$.
    \protect\end{enumerate}
    Here, we set $p_{0}=0.65$ and $P_M/P_H=1.2$, and parameters $\alpha$, $\mu$, $\gamma$, $\nu$, $g=\omega(t)$, $\beta$, $p_{m \to h}$ as per Table \ref{table:model_params}.}
    \label{fig:imm_short_latency_ss}
\end{figure}

We now perform a sensitivity analysis for endemic equilibrium solutions allowing for the acquisition of transmission-blocking and antidisease immunity (Figure \ref{fig:imm_short_latency_ss}). Here, we restrict ourselves to short-latency strains ($k=0$). For a fixed set of hypnozoite activation and death rates $(\alpha, \mu)$, long-latency strains ($k>0$) yield similar qualitative patterns as a function of the immunity parameters $w$, $p_{tb}$ and $p_c$. In the absence of transmission-blocking and clinical immunity (that is, $p_{tb}=p_c=1$), we revert to the setting examined in Section \ref{sec::sens_no_imm}; we highlight this case with closed circles in Figure \ref{fig:imm_short_latency_ss}.\\

A two-way sensitivity analysis, with respect to parameters $w$ and $(1-p_{tb})$, is shown in Figure \ref{fig:imm_short_latency_ss}A. Recall that the probability of human-to-mosquito transmission (when an uninfected mosquito takes a  bloodmeal from a blood-stage infected human) decays geometrically, with factor $p_{tb}$, as a function of an individual's immunity level (Section \ref{sec::tb_imm}); since protection rises as $p_{tb} \to 0$, we think of $(1-p_{tb})$ as a transmission-blocking protection parameter.
In contrast, the parameter $1/w$ governs the time scale for which immunity is retained, with the limiting case $w = 0$ corresponding to a scenario where immunity is never lost.  The endemic equilibrium prevalence of blood-stage infection (Figure \ref{fig:imm_short_latency_ss}A1) decreases both as:
\begin{itemize}
    \item immunity becomes longer-lived (that is, $w \to 0$), whereby a larger subset of an individual's infection history is expected to contribute to their current immunity level; and
    \item the protective effect associated with each cleared blood-stage infection is augmented (that is, $(1-p_{tb}) \to 1$).
\end{itemize}
Mitigation of the blood-stage infection burden in light of transmission-blocking immunity, however, necessarily limits exposure; reductions in the the prevalence of blood-stage infection are therefore accompanied by reductions in the population-level distribution of immunity. The mean immunity level at the endemic equilibrium therefore \textit{decreases}, even as the rate of immune decay $w$ decreases, and immunity accrues over a larger time scale (Figure \ref{fig:imm_short_latency_ss}A2). Likewise, augmenting the transmission-blocking effect of each immunity increment $(1-p_{tb})$ --- whereby the probability of human-to-mosquito is suppressed strongly, even at low immunity levels --- leads to a reduction in the mean immunity level at the endemic equilibrium (Figure \ref{fig:imm_short_latency_ss}A2).\\

In particular, we see a substantially reduced burden of (blood-stage) superinfection at the endemic equilibrium as the transmission-blocking effect of each immune increment $(1-p_{tb})$ increases (Figure \ref{fig:imm_short_latency_ss}B). The suppression of superinfection explains the stronger decay in the mean immunity level (Figure \ref{fig:imm_short_latency_ss}A2), relative to the prevalence of blood-stage infection (Figure \ref{fig:imm_short_latency_ss}A1), as the transmission-blocking protection parameter $(1-p_{tb})$ is strengthened: since the clearance of each primary infection and relapse yields an immunity boost, irrespective of temporal overlap with other blood-stage infections, superinfection is an important driver of acquired immunity.\\

We also observe a trade-off between transmission-blocking and antidisease immunity. The mitigation of transmission as the transmission-blocking protection parameter $(1-p_{tb})$ is augmented leads to a lower population-level distribution of immunity at the endemic equilibrium (Figure \ref{fig:imm_short_latency_ss}A2). If the distribution of immunity at the endemic equilibrium is sufficiently reduced, then an increasingly strong transmission-blocking effect per immune increment $(1-p_{tb})$ can give rise to an increasing prevalence of clinical infection at the endemic equilibrium (Figure \ref{fig:imm_short_latency_ss}C), even as the burden of blood-stage infection continues to decline (Figure \ref{fig:imm_short_latency_ss}B).

\subsection{A reduced system of IDEs to study transient behaviour} \label{sec::hybrid_ide}

The hybrid transmission model given by Equations (\ref{hybrid_human_ode}) to (\ref{hybrid_mos_ode}) yields population-level dynamics of superinfection, the hypnozoite reservoir and acquired  immunity. However, the countable system of ODEs given by Equations (\ref{hybrid_human_ode}) to (\ref{hybrid_mos_ode}) is not necessarily readily amenable to numerical solution; truncating the system at reasonable endpoints could yield thousands of coupled ODEs, since the size of the hypnozoite reservoir (and, by extension, the immunity level) could reasonably be expected to range up to 30 in moderate to high transmission settings (see Figure 4 of \textcite{white2014modelling} and Figure 6 of \textcite{anwar2021multiscale}).\\

Here, we propose a reduced system of integrodifferential equations (IDEs) to couple host and vector dynamics, following the approach detailed in \textcite{mehra2022superinf}. In particular, we observe that:
\begin{itemize}
    \item The dependence of the human population on vector dynamics can be distilled into the FORI, which is proportional to the number of infected mosquitoes in the population $I_M(t)$ (see Equation (\ref{fori})).
    \item The dependence of the vector population on the state of the human population can be distilled into the probability of successful human-to-mosquito transmission $p_{h \to m}(t)$ when an uninfected mosquito bites \textit{any} human in the population; note that the quantity $p_{h \to m}(t)$ accounts for \textit{both} the prevalence of blood-stage infection and the distribution of (transmission-blocking) immunity within the human population.
\end{itemize}

At time $t=0$, we make the assumption that each individual in the human population (of fixed size $P_H$) has immunity level zero; carries no hypnozoites; and harbours no ongoing blood-stage infections, whereby $p_{h \to m}(0) =0 $. As such, we consider the introduction of a number of infected mosquitoes into an otherwise infection- and immune-naive human population. As a function of the FORI, the probability of successful human-to-mosquito transmission $p_{h \to m}(t)$ is then governed by the integral given in Equation (\ref{prob_h_to_m}). Likewise, as a function of the probability of successful human-to-mosquito transmission $p_{h \to m}(t)$ per bloodmeal, the time evolution of the expected number of infected mosquitoes over time $I_M(t)$ is governed by the coupled system of ODEs given by Equations (\ref{hybrid_inf_mos_ode}) to (\ref{hybrid_mos_ode}), which also captures the time evolution of the expected number of latent $L_M(t)$ and uninfected $U_M(t)$ mosquitoes over time.\\

Coupling expected host and vector dynamics under a hybrid approximation thus yields the system of IDEs
\begin{align}
    \frac{dI_M}{dt} &= \eta L_M(t) - gI_M(t) \label{eq:hybrid_im} \\
    \frac{dL_M}{dt} &= \beta(t) p_{h \to m}(t) U_M(t) - (g + \eta) L_M(t) \\
    \frac{dU_M}{dt} &= \omega(t) \big[ I_M(t) + L_M(t) + U_M(t) \big] - \big[ g + \beta(t) p_{h \to m}(t) \big] U_M(t) \label{eq:hybrid_um}\\
    p_{h \to m}(t) &= p_0 \Bigg( \exp \bigg\{ - \int^t_0 \beta(\tau) p_{m \to h} \frac{I_M(\tau)}{P_H} \bigg[ 1 -  \frac{1 - (1-p) p_{h, I}(t - \tau)} {1 + \nu (1-p) p_{p, I}(t-\tau)} \bigg] d \tau \bigg\} \notag \\
    &  \qquad - \exp \bigg\{ - \int^t_0 \beta(\tau) p_{m \to h} \frac{I_M(\tau)}{P_H} \bigg[ 1 -  \frac{1 - (1-p) \cdot p_{p, I}(t-\tau) - p_{p, A}(t-\tau) } {1 + \nu (1-p) p_{h, I}(t-\tau) + \nu p_{h, A}(t-\tau)} d \tau \bigg\} \Bigg) \label{eq:hybrid_phm}
\end{align}
with initial condition $I_M(0), L_M(0), U_M(0) \geq 0$, where we have used Equations (\ref{prob_h_to_m}) and (\ref{hybrid_inf_mos_ode}) to (\ref{hybrid_mos_ode}). Recall that 
\begin{itemize}
    \item $p_{h, A}(x)$ denotes the probability that a hypnozoite has activated to give rise to a relapse that is ongoing time $x$ after inoculation (Equation (\ref{a_eq})).
    \item $p_{h, I}(x)$ denotes the probability that immune memory has been gained (following the clearance of a relapse) time $x$ after a hypnozoite is established in the liver (Equation (\ref{i_eq})).
    \item $p_{p, A}(x)$ denotes the probability that a primary infection is ongoing time $x$ after onset (Equation (\ref{p_eq})).
    \item $p_{p, I}(x)$ denotes the probability that immune memory has been gained time $x$ after the onset of a primary infection (Equation (\ref{p_eq})).
\end{itemize}
Interpretations for each transmission/within-host parameter are detailed in Table \ref{table:model_params}.\\

The system of IDEs given by Equations (\ref{eq:hybrid_im}) to (\ref{eq:hybrid_phm}) couples expected host and vector dynamics, whilst concurrently capturing the accrual of the hypnozoite reservoir (either short- or long-latency strains); superinfection; transmission-blocking immunity and fluctuations in the mosquito population size (due to seasonality or the implementation of vector-based control measures).\\

Observe that the integral equation governing $p_{h \to m}(t)$ (Equation (\ref{eq:hybrid_phm})) --- which we derived using the within-host PGF given by Equation (\ref{vivax_multi_pgf_imm}) --- satisfies the Kolmogorov forward differential equations for the within-host queueing structure, which in turn constitutes the expected frequency distribution of the human population (Equation (\ref{hybrid_human_ode})), granted an individual is both immune- and infection-naive at time zero and the FORI at time $t$ is $\beta(t) p_{m \to h} I_M(t)/P_H$. As such, under the assumption that the human population is both immune- and infection-naive at time zero, the time evolution of the FORI and the probability of human-to-mosquito transmission per bloodmeal are equivalent under the system of IDEs given by Equations (\ref{eq:hybrid_im}) to (\ref{eq:hybrid_phm}), and the countable system of ODEs given by Equations (\ref{hybrid_human_ode}) to (\ref{hybrid_mos_ode}).\\

While the quantities $I_M(t)$, $L_M(t)$, $U_M(t)$ and $p_{h \to m}(t)$ are sufficient to couple host and vector dynamics, we ultimately seek to characterise population-level distributions for quantities of epidemiological interest. We note, however, that the complete population-level distribution of superinfection, immunity and hypnozoite states can be recovered conditional on the FORI using the results derived in \textcite{mehra2022hypnozoite}. To reiterate the premise of the hybrid approximation, the population-level transmission models discussed here have been constructed by casting the within-host probabilistic distribution as the population-level frequency distribution \parencite{henry2020hybrid}. Given the time evolution of the FORI $\lambda(t)=\beta(t) p_{m \to h} I_M(t)/P_H$ derived from the system of IDEs given by Equations (\ref{eq:hybrid_im}) to (\ref{eq:hybrid_phm}), we can recover population-level distributions for quantities of epidemiological interest using the formulae derived in \textcite{mehra2022hypnozoite}.\\

The methodology adopted here uses an integral system, within which we can enforce time-dependence in the bite rate per mosquito $\beta(t)$ and the mosquito birth rate $\omega(t)$. In Section \ref{appendix:illustrative_ide} below, we present illustrative results for two scenarios: a constant transmission setting, where all transmission parameters are fixed (Section \ref{sec::ide_const_transmission_setting}); and a seasonal transmission setting, with a sinusoidal mosquito birth rate $\omega(t)$ (Section \ref{sec::ide_seasonal_transmission_setting}). Vector-based control interventions represent a natural extension \parencite{le2007elaborated, griffin2010reducing, white2018mathematical}, but are not presented here.

\subsubsection{Illustrative results for the reduced system of IDEs} \label{appendix:illustrative_ide}

To recover transient host and vector dynamics, we solve the system of IDEs given by Equations (\ref{eq:hybrid_im}) and (\ref{eq:hybrid_phm}) numerically, using Euler's method (for the ODEs given by Equations (\ref{eq:hybrid_im}) to (\ref{eq:hybrid_um})) and the trapezoidal rule (for the integral given by Equation (\ref{eq:hybrid_phm})) with a fixed time step; this procedure is a variation of the algorithm proposed by \textcite{anwar2021multiscale}. As a function of the FORI derived from Equations (\ref{eq:hybrid_im}) and (\ref{eq:hybrid_phm}), we recover the time evolution of several quantities of epidemiological interest. Relevant formulae (as per \textcite{mehra2022hypnozoite}) are provided in Appendix \ref{appendix::steady_state_dist}, including:
\begin{itemize}
    \item the mean and variance for the size of the (non)-latent hypnozoite reservoir (Equations (\ref{exp_hyp}) and (\ref{var_hyp}));
    \item the PMF for the number of co-circulating blood-stage broods (Equations (\ref{moi_0}) to (\ref{moi_2}));
    \item the relapse rate conditional on the blood-stage infection status (Equations (\ref{rel_rate_moi_0}) to (\ref{rel_rate_moi_3+}));
    \item the distribution of immunity, as quantified by the mean and variance of the discrete immunity levels (Equations (\ref{exp_imm}) and (\ref{var_imm}));
    \item the prevalence of clinical infection (Equation (\ref{prob_clin})).
\end{itemize}

\paragraph{Non-seasonal transmission} \label{sec::ide_const_transmission_setting}

\begin{figure}
    \centering
    \includegraphics[width=0.94\textwidth]{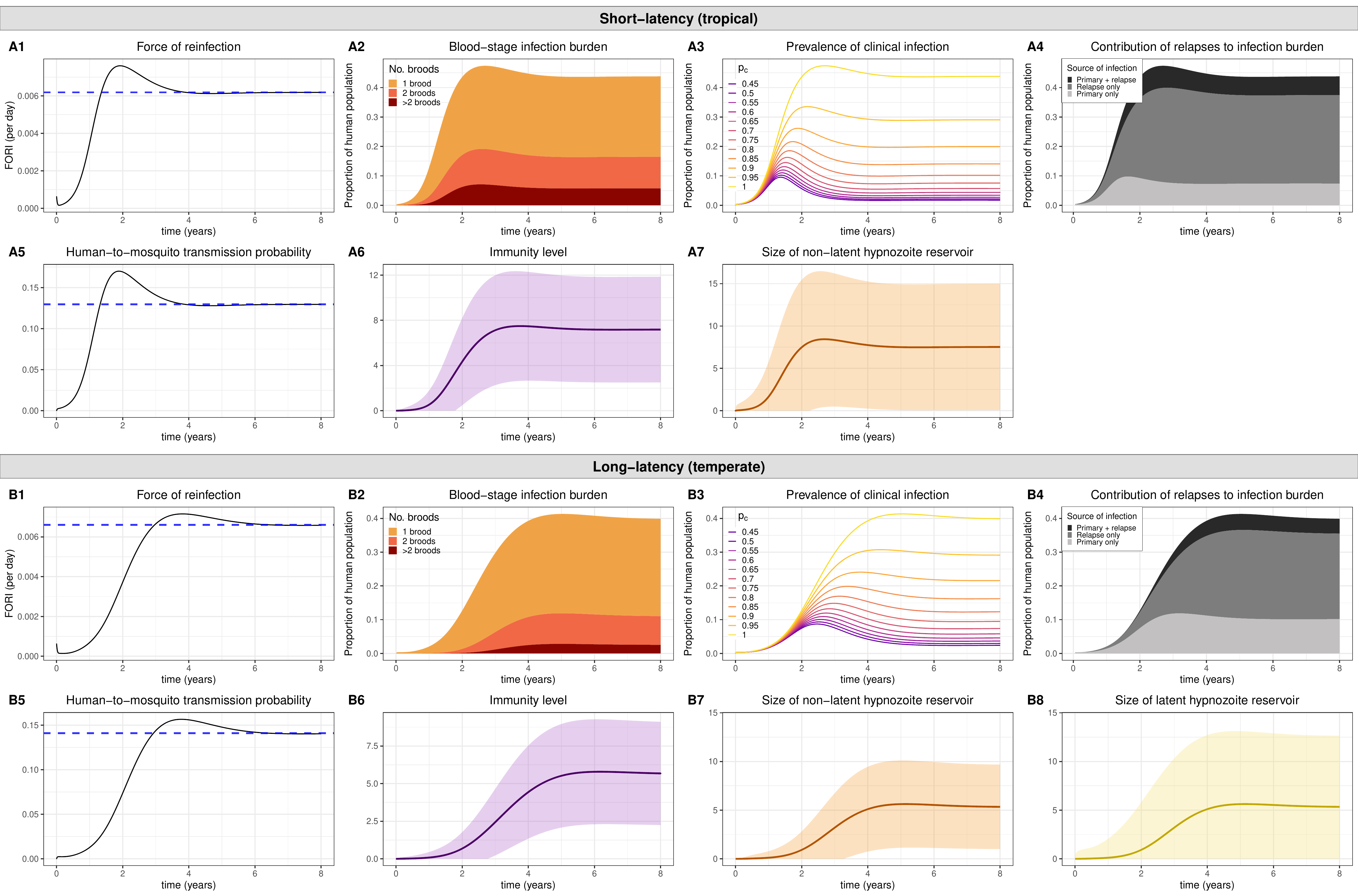}
    \captionsetup{singlelinecheck=off}
    \caption{Transient host and vector dynamics for (A) short-latency ($k=0$) and (B) long-latency ($k=2$, $\delta=1/100$ day$^{-1}$) strains. Here, we consider a constant transmission setting with $\beta(t) = \beta$ and $\omega(t) = g$. At time zero, we assume that the human population is both infection- and immune-naive, with $I_M(0)/P_H = 0.012, L_M(0)/P_H = 0, U_M(0)/P_H = 1.2$. We numerically solve the system over a period of $8$ years, with a fixed time step of $0.1$ days. The (A1/B1) FORI $\beta p_{m \to h} \frac{I_M(t)}{P_H}$ and (A5/B5) probability of human-to-mosquito tranmission $p_{h \to m}(t)$ are governed by the system of IDEs given by Equations (\ref{eq:hybrid_im}) and (\ref{eq:hybrid_phm}). Endemic equilibrium solutions for the FORI $\beta p_{m \to h} \frac{I_M*}{P_H}$ and human-to-mosquito transmission probability $p^*_{h \to m}$, given by the non-trivial solution to Equations (\ref{im_sim_eq_1}) and (\ref{im_sim_eq_2}), are indicated with dashed blue lines. All other quantities are calculated as a function of the numerical solution for the FORI $\beta p_{m \to h} \frac{I_M(t)}{P_H}$, including
    \protect\begin{itemize}
        \protect\item (A2/B2): the PMF for the number co-circulating parasite broods (Equations (\ref{moi_0}) to (\ref{moi_2}))
        \protect\item (A3/B3): the prevalence of clinical infection (Equation (\ref{prob_clin})), with $p_c$ ranging from $0.45$ to $1$ in increments of $0.05$
        \protect\item (A4/B4): the respective contributions of relapses and primary infections to the burden of blood-stage infection
        \protect\item (A6/B6): the mean immunity level (Equation (\ref{exp_imm})) (shading indicates one standard deviation (Equation (\ref{var_imm})) above and below the mean)
        \protect\item (A7/B7, A8/B8): the expected size of the (non)-latent hypnozoite reservoir (Equation (\ref{exp_hyp})) (shading indicates one standard deviation (Equation (\ref{var_hyp})) above and below the mean)
    \protect\end{itemize}
    Model parameters $\alpha, \mu, \gamma, \nu, w, g, \eta, \beta, p_{m \to h}, p_{tb}$ as per Table \ref{table:model_params}, with $p_0 = 0.65$.}
    \label{fig:illustrative_ide}
\end{figure}

We begin by considering host and vector dynamics in the absence of seasonality (Figure \ref{fig:illustrative_ide}). At time zero, we consider the introduction of several infected mosquitoes into an (blood- and liver-stage) infection and immune naive human population. Predicted endemic equilibrium solutions, given by the unique non-trivial solution to Equations (\ref{im_sim_eq_1}) and (\ref{im_sim_eq_2}), are shown with dashed blue lines for the FORI (Figures \ref{fig:illustrative_ide}A1/B1) and the immunity-modulated probability of human-to-mosquito transmission (Figures \ref{fig:illustrative_ide}A5/B5).\\

Illustrative dynamics for short latency strains ($k=0$) are shown in Figure \ref{fig:illustrative_ide}A. The low initial level of infection in the mosquito population constrains the transmission intensity at early time points. Prior to the acquisition of extensive transmission-blocking immunity --- with relatively low immunity levels harboured for a year following the introduction of infected mosquitoes into an immune-naive human population (Figure \ref{fig:illustrative_ide}A6) --- human-to-mosquito transmission remains comparatively unconstrained, leading to a sustained increase in the FORI (Figure \ref{fig:illustrative_ide}A1), and consequently, the blood-stage infection burden  (Figure \ref{fig:illustrative_ide}A2) and the size of the hypnozoite reservoir (Figure \ref{fig:illustrative_ide}A7). A pronounced rise in the prevalence of blood-stage infection during this early period leads to an increase in both the probability of human-to-mosquito transmission (Figure \ref{fig:illustrative_ide}A5) and the prevalence of clinical infection (Figure \ref{fig:illustrative_ide}A3). Intensified transmission, however, is accompanied by the sustained acquisition of immunity (Figures \ref{fig:illustrative_ide}A6), which eventually mitigates the probability of human-to-mosquito transmission (Figure \ref{fig:illustrative_ide}A5), leading to a reduction in the FORI (Figure \ref{fig:illustrative_ide}A5), as well as a slight reduction in the burden of (clinical) blood-stage infection (Figures \ref{fig:illustrative_ide}A2, A3). These transient effects eventually subside, and for the set of parameters considered here, the predicted endemic equilibrium (obtained by numerically solving Equations (\ref{im_sim_eq_2}) and (\ref{im_sim_eq_1}), and indicated with dotted lines blue lines) is reached within approximately $4$ years.\\

Analogous results for long-latency strains ($k>0$) are shown in Figure \ref{fig:illustrative_ide}B. As a consequence of the enforced hypnozoite dormancy period (with expected duration $k/\delta = 200$ days and standard deviation $\sqrt{k}/\delta = 100 \sqrt{2}$ days), the non-latent hypnozoite reservoir remains limited in size for approximately one year (Figure \ref{fig:illustrative_ide}B7); as such, single-brood primary infections dominate the infection burden for an extended period of time (Figures \ref{fig:illustrative_ide}B2 and \ref{fig:illustrative_ide}B4) relative to short-latency strains. In tandem with the emergence of hypnozoites from dormancy, relapses eventually drive up the burden of (clinical) blood-stage infection (Figures \ref{fig:illustrative_ide}B2 and \ref{fig:illustrative_ide}B4). As for short-latency strains ($k=0$), the acquisition of transmission-blocking immunity eventually mitigates onward human-to-mosquito transmission (Figure \ref{fig:illustrative_ide}B5), leading to a slight reduction in transmission intensity before the predicted endemic equilibrium (obtained by numerically solving Equations (\ref{im_sim_eq_2}) and (\ref{im_sim_eq_1}), and indicated with dotted lines blue lines) is reached six years after the introduction of infected mosquitoes into an infection- and immune-naive human population.

\paragraph{Seasonal transmission} \label{sec::ide_seasonal_transmission_setting}

\begin{figure}
    \centering
    \includegraphics[width=0.94\textwidth]{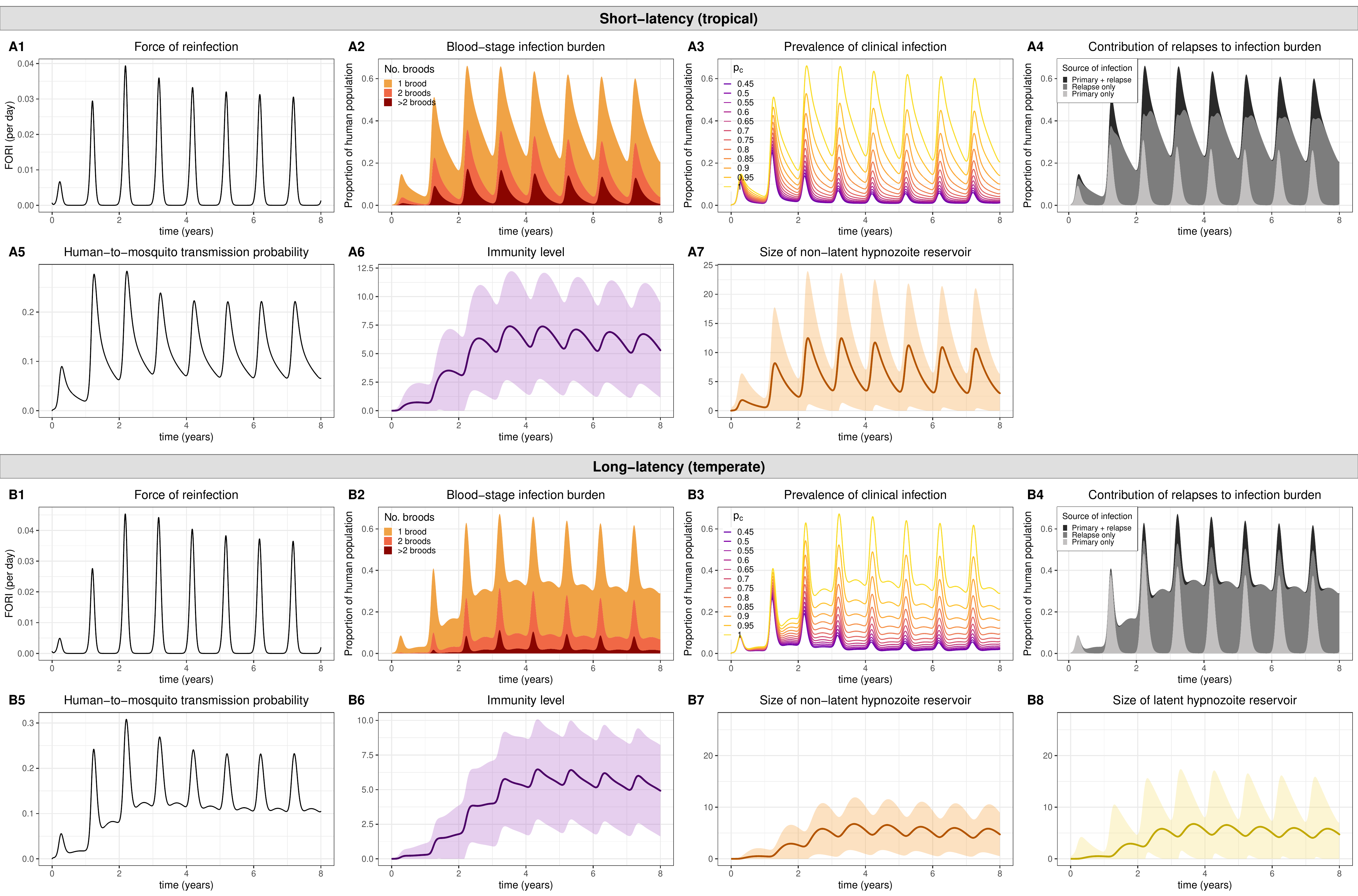}
    \captionsetup{singlelinecheck=off}
    \caption{Transient host and vector dynamics for (A) short-latency ($k=0$) and (B) long-latency ($k=2$, $\delta=1/100$ day$^{-1}$) strains. Here, we consider a seasonal transmission setting with $\beta(t) = \beta$ and $\omega(t) = g \big[ \sin(\frac{2 \pi t}{365} + \frac{3 \pi}{4}) + 1 \big]$. At time zero, we assume that the human population is both infection- and immune-naive, with $I_M(0)/P_H = 0.012, L_M(0)/P_H = 0, U_M(0)/P_H = 1.2$. We numerically solve the system over a period of $8$ years, with a fixed time step of $0.02$ days. The (A1/B1) FORI $\beta p_{m \to h} \frac{I_M(t)}{P_H}$ and (A5/B5) probability of human-to-mosquito tranmission $p_{h \to m}(t)$ are governed by the system of IDEs given by Equations (\ref{eq:hybrid_im}) and (\ref{eq:hybrid_phm}). All other quantities are calculated as a function of the numerical solution for the FORI $\beta p_{m \to h} \frac{I_M(t)}{P_H}$, including
    \protect\begin{itemize}
        \protect\item (A2/B2): the PMF for the number co-circulating parasite broods (Equations (\ref{moi_0}) to (\ref{moi_2}))
        \protect\item (A3/B3): the prevalence of clinical infection (Equation (\ref{prob_clin})), with $p_c$ ranging from $0.45$ to $1$ in increments of $0.05$
        \protect\item (A4/B4): the respective contributions of relapses and primary infections to the burden of blood-stage infection
        \protect\item (A6/B6): the mean immunity level (Equation (\ref{exp_imm})) (sharing indicates one standard deviation (Equation (\ref{var_imm})) above and below the mean)
        \protect\item (A7/B7, A8/B8): the expected size of the (non)-latent hypnozoite reservoir (Equation (\ref{exp_hyp})) (shading indicates one standard deviation (Equation (\ref{var_hyp})) above and below the mean)
    \protect\end{itemize}
    Model parameters $\alpha, \mu, \gamma, \nu, w, g, \eta, \beta, p_{m \to h}, p_{tb}$ as per Table \ref{table:model_params}, with $p_0 = 0.65$.}
    \label{fig:illustrative_ide_seasonal}
\end{figure}

To allow for seasonality, arising, for instance, from external climactic variation, we impose sinusoidal forcing (with period one year) on the mosquito birth rate. Illustrative dynamics for short-latency ($k=0$) and long-latency ($k>0$) strains are shown in Figures \ref{fig:illustrative_ide_seasonal}A and \ref{fig:illustrative_ide_seasonal}B respectively. With the imposition of seasonal forcing, infection levels within both humans and mosquitoes exhibit oscillations (with period one year) that eventually stabilise around a steady mean. The nature of these oscillations within a season, however, varies between short- and long-latency strains. Oscillations in the FORI (Figure \ref{fig:illustrative_ide_seasonal}A1/B1) are driven strongly by seasonal fluctuations in the size of the mosquito population. For short-latency strains ($k=0$), the burden of (clinical) blood-stage infection decays monotonically between seasonal peaks (Figures \ref{fig:illustrative_ide_seasonal}A2 and \ref{fig:illustrative_ide_seasonal}A3) as the hypnozoite reservoir is depleted in light of limited mosquito-to-human transmission (as quantified by the FORI, Figure \ref{fig:illustrative_ide_seasonal}A1). For long-latency strains, yearly maxima in the burden of (clinical) blood-stage infection (Figures \ref{fig:illustrative_ide_seasonal}B2 and \ref{fig:illustrative_ide_seasonal}B3) likewise coincide with seasonal peaks in the FORI (Figure \ref{fig:illustrative_ide_seasonal}B1), with primary infections contributing to the majority of the blood-stage infection burden during these periods of intensified mosquito-to-human transmission (Figure \ref{fig:illustrative_ide_seasonal}B4). As a consequence of the hypnozoite dormancy period --- which introduces a delay between periods of intensified mosquito-to-human transmission and elevated relapse risk (as quantified through the size of the non-latent hypnozoite reservoir, Figure \ref{fig:illustrative_ide_seasonal}B7) --- we observe biphasic infection dynamics, whereby the burden of (clinical) blood-stage infection exhibits a second, smaller peak approximately 6 months after the yearly maximum (Figures \ref{fig:illustrative_ide_seasonal}B2 and \ref{fig:illustrative_ide_seasonal}B3) driven by relapses (Figures \ref{fig:illustrative_ide_seasonal}B4) as hypnozoites established during the seasonal peak of mosquito-to-human transmission emerge from dormancy. As such, we predict that hypnozoite dormancy sustains the burden of blood-stage infection between seasonal peaks in the FORI, in line with the hypothesis that long-latency phenotypes evolved in temperate regions to sustain transmission despite of limited mosquito breeding during the winter season \parencite{white2016some}.

\section{Discussion} \label{sec::discussion}

The interplay between the hypnozoite reservoir, superinfection and acquired immunity is a key aspect of the epidemiology of \textit{P. vivax}. Here, we propose a hybrid transmission model for \textit{P. vivax}, accounting for hypnozoite accrual, (blood-stage) superinfection and the acquisition of transmission-blocking and antidisease immunity. To capture within-host dynamics as a function of the FORI, we extend the open network of infinite server queues constructed in \textcite{mehra2022hypnozoite} to embed a discretised version of the antibody model we introduced in \textcite{mehra2021antibody}. By deriving the joint PGF for the state of the queueing network (Equation (\ref{vivax_multi_pgf_imm})), we obtain an analytic description of within-host dynamics in a general transmission setting. To couple host and vector dynamics, we adopt the hybrid approximation of \textcite{nasell2013hybrid} under which probabilistic within-host distributions are cast as expected population-level proportions \parencite{henry2020hybrid}. We thus recover a deterministic compartmental model, comprising a countably infinite system of ODEs (Equations (\ref{hybrid_human_ode}) and (\ref{hybrid_mos_ode})), which can be viewed as a natural extension of the Ross-Macdonald framework. For a simpler system with countably many states, we demonstrated the equivalence of the hybrid approximation to the functional law of large numbers \parencite{barbour2012law} for an appropriate Markov population process in \textcite{mehra2022superinf}.\\

We draw on distributions derived at the within-host level \parencite{mehra2022hypnozoite} to characterise both the transient and steady state behaviour of this compartmental model. In particular, following the approach we developed in \textcite{mehra2022superinf}, we derive a reduced system of IDEs governing the time evolution of the number of (un)infected and latent mosquitoes; and the immunity-modulated probability of human-to-mosquito transmission (Equations (\ref{eq:hybrid_im}) to (\ref{eq:hybrid_phm})). As a function of the FORI predicted under this reduced system of IDEs --- which is equivalent to the complete compartmental model granted the human population is initially immune- and infection-naive --- we recover complete population-level distributions for various quantities of epidemiological interest, using the formulae derived in \textcite{mehra2022hypnozoite} (see Appendix \ref{appendix::steady_state_dist}). By drawing on the within-host queueing models we introduced in \textcite{mehra2021antibody, mehra2022hypnozoite}, and the construction developed in \textcite{mehra2022superinf}, we circumvent the practical constraints that have previously limited the tractability of hypnozoite density models \parencite{white2018mathematical}.\\

Our model, to the best of our knowledge, provides the most complete description of superinfection, immunity and hypnozoite dynamics for \textit{P. vivax} thus far, while remaining readily amenable to numerical solution and analysis. In \textcite{mehra2022superinf}, we developed a framework to capture the dynamics of (short-latency) hypnozoite accrual and superinfection, addressing a gap in the literature with respect to the rigorous analysis of superinfection; we have extended the framework of \textcite{mehra2022superinf} in the present manuscript to allow for greater biological realism, namely, acquired immunity and long-latency phenotypes. The joint population-level dynamics of the hypnozoite reservoir and acquired immunity have been previously examined by \textcite{white2018mathematical}. The construction of \textcite{white2018mathematical} is predicated on a continuous age- and exposure-dependent immunity level, which is subsequently mapped (using Hill functions) to correlates of antidisease immunity (that is, a reduced probability of clinical infection) and antiparasite immunity (including accelerated parasite clearance and mitigated parasite densities, manifesting in a reduced probability of detection via light microscopy). Here, we instead consider a discretised exposure-dependent immunity level, which we map to correlates of antidisease immunity and transmission-blocking immunity (that is, a reduced probability of human-to-mosquito transmission) under the assumption of geometric decay. While \textcite{white2018mathematical} explicitly account for treatment, and age structure and heterogeneity in the human population, we restrict our attention to a homogeneous human population in the absence of treatment and age structure. Unlike \textcite{white2018mathematical}, however, we monitor hypnozoite densities rather than broods (thereby capturing variation in parasite inocula across bites), in addition to the population-level distribution of superinfection; our framework also holds for long-latency hypnozoite strains, unlike the framework of \textcite{white2018mathematical} which is restricted to short-latency strains.\\

While most previous hypnozoite `batch' and `density' models \parencite{white2014modelling, white2018mathematical, anwar2021multiscale} have relied on numerical solution to characterise steady state solutions, our sensitivity analyeses are informed by the within-host distributions derived in \textcite{mehra2022hypnozoite}. We recover a threshold phenomenon for the hybrid model, deriving a bifurcation parameter (Equation (\ref{eq:R0})) governing the existence of endemic equilibria. In the absence of transmission-blocking immunity ($p_{tb} = 1$) and mosquito latency ($1/\eta = 0$), we were able to perform an asymptotic stability analysis in \textcite{mehra2022superinf} for the first-order IDE governing the time-evolution of the FORI using the stability criterion of \textcite{brauer1978asymptotic}; the imposition of transmission blocking immunity ($p_{tb} < 1$) or mosquito latency ($1/\eta > 0$) yields a higher-order system of IDEs governing the FORI, for which we are unaware of asymptotic stability criteria.\\

The transient and stationary results presented in this manuscript are underpinned by analyticity at the within-host level, which, in turn, is predicated on the assumption that each hypnozoite/infection is governed by an independent stochastic process. The assumption of independent, spontaneous hypnozoite activation in line with the `genetic clock' hypothesis, as implemented by \textcite{white2014modelling}, is critical to our construction:  external triggers of hypnozoite activation (e.g. febrile illness, arising from parasitic or bacterial infections \parencite{shanks2013activation}, particularly \textit{P. falciparum} \parencite{commons2019risk}) necessarily introduce synchronicity between activating hypnozoites, thereby violating the assumption of independent hypnozoite behaviour. Our model does not readily accommodate interactions between concurrent hypnozoites/infections, for instance, competition between co-circulating parasite broods \parencite{de2005virulence}. Antiparasite immunity (manifest in the modulation parasite clearance rates \parencite{white2017malaria}) and pre-erythrocytic immunity \parencite{mueller2013natural}, which render hypnozoite/infection dynamics dependent on the infection history, are likewise intractable. On a population-level, our model is constrained by the assumption of homogeneity for the human population. We do not account for age structure or demography within the human population; heterogeneity in the risk of relapse and immunity levels across individuals is driven purely by stochastic fluctuations, rather than differences in the time over which the hypnozoite reservoir has been accrued and immunity has been acquired.\\

Our formulation of immunity, moreover, is non-mechanistic and subject to a number of simplifying assumptions. Adopting a discretised version of the model presented in \textcite{mehra2021antibody}, we assume that the clearance of each blood-stage infection is accompanied by an immunity boost with unit magnitude and an exponentially-distributed lifetime. Empirical characterisation of antibody titres, however, has revealed substantial heterogeneity in the magnitude of antibody boosts across successive infections, and the temporal distribution of antibody boosts associated with different antigens \parencite{white2014dynamics}. A key omission in our model is strain specificity; while homologous challenge yields a strong immune response, immune protection following heterologous challenge is contingent on the degree of cross-reactivity between strains \parencite{mueller2013natural}. As such, the discretised immunity level considered here largely serves as a proxy for `recent' exposure to blood-stage infection, with the immune decay parameter $w$ governing the time scale on which immunity is retained.\\

Nonetheless, in capturing the interplay between hypnozoite accrual, superinfection and acquired immunity --- and providing, to the best of our knowledge, the most complete population-level distributions for a range of epidemiological values --- our model provides insights into important, but poorly understood, epidemiological features of \textit{P. vivax}, with natural extensions to explore the consequences of control and elimination strategies.

\section*{Acknowledgements}
Jennifer A. Flegg's research is supported by the Australian Research Council (ARC) (DE160100227 and DP200100747). Peter G. Taylor's research is supported by the ARC Centre of Excellence for Mathematical and Statistical Frontiers (ACEMS) (CE140100049). James M. McCaw's research is supported by the ARC (DP210101920 and ARC DP170103076).

\newpage

\appendix

\section{State probabilities for a single hypnozoite} \label{sec::single_hyp}
Here, we provide solutions to the system of ODEs given by Equations (\ref{hyp1_Kolmogorov}) to (\ref{hypI_Kolmogorov}). We have previously solved related models in \textcite{mehra2020activation, mehra2022hypnozoite}: 
\begin{itemize}
    \item In \textcite{mehra2020activation}, we analysed the activation-clearance model proposed by \textcite{white2014modelling}, but did not model the dynamics of blood-stage infection following each hypnozoite activation event (that is, states $A$, $I$ and $C$ were not distinguished).
    \item In \textcite{mehra2022hypnozoite}, we constructed a relapse-clearance model, extending the model proposed in \textcite{white2014modelling} (and solved in \textcite{mehra2020activation}) to accommodate a blood-stage infection (of exponentially-distributed duration) following hypnozoite activation, but did not capture the acquisition/waning of immunity (that is, states $I$ and $C$ were not distinguished).
\end{itemize}

As in \textcite{mehra2020activation} and \textcite{mehra2022hypnozoite}, we solve the system of ODEs successively using integration by parts, drawing on standard integral number 2.321.2 of \textcite{jeffrey2007table}, to give:

\begin{align}
	p_{h,m}(t) =& \frac{(\delta t)^{m-1}}{(m-1)!}e^{-(\mu+\delta)t} \text{ for } m \in [1, k] \label{l_eq}\\
	p_{h,NL}(t) =& \frac{\delta^{k}}{(\delta - \alpha)^{k}} \Bigg[ e^{-(\mu + \alpha)t} - e^{-(\mu + \delta)t} \sum^{k-1}_{j=0} \frac{t^j}{ j!} (\delta - \alpha)^{j} \Bigg] \label{nl_eq} \\
    p_{h,A}(t) =& \frac{\alpha \delta^{k}}{(\delta - \alpha)^{k}} \Bigg[\frac{e^{-(\mu + \delta)t}}{\mu + \delta - \gamma} \Bigg\{ \sum^{k-1}_{i=0} t^i \bigg[ \frac{(\mu -\gamma + \delta)^{i}}{i!} \sum^{k-1}_{j=i} \Big( \frac{\delta - \alpha}{\mu - \gamma + \delta} \Big)^j \bigg]  \Bigg\} - \frac{e^{-(\mu + \alpha)t}}{\mu - \gamma + \alpha} \Bigg] + \notag \\
    & \frac{\alpha}{\alpha + \mu - \gamma} \Big( \frac{\delta}{\delta + \mu - \gamma} \Big)^{k} e^{-\gamma t} \label{a_eq} \\
    p_{h,I}(t) =& \frac{\alpha}{\alpha + \mu - \gamma} \Big( \frac{\delta}{\delta - \alpha} \Big)^k \frac{\gamma}{\alpha + \mu -w } e^{-(\mu + \alpha)t} + \frac{\gamma}{w - \gamma} \frac{\alpha}{\alpha + \mu - \gamma} \Big( \frac{\delta}{\delta + \mu - \gamma} \Big)^{k} e^{-\gamma t} \notag\\
    & - \frac{\gamma \alpha \delta^k}{(\delta - \alpha)^k} \frac{e^{-(\mu + \delta) t}}{(\mu + \delta - \gamma)(\mu + \delta - w)} \Bigg\{ \sum^{k-1}_{\ell=0} t^\ell \Bigg[ \frac{(\mu + \delta - w)^{\ell}}{\ell!} \sum^{k-1}_{i=\ell} \Big( \frac{\mu+\delta-\gamma}{\mu+\delta-w} \Big)^i \sum^{k-1}_{j=i} \Big( \frac{\delta-\alpha}{\mu + \delta - \gamma}\Big)^j \Bigg\} \Bigg] \notag\\
    & + \frac{\gamma}{\gamma-w} \frac{\alpha}{\alpha + \mu - w} \Big( \frac{\delta}{\delta + \mu - w} \Big)^k \label{i_eq}
\end{align}

Solutions to $p_{h,m}(t)$, $m\in [1, k]$ and $p_{h,NL}(t)$ were intially presented in \textcite{mehra2020activation}, while solutions to $p_{h,A}(t)$ were initially derived in \textcite{mehra2022hypnozoite}.

\section{Population-level distributions for quantities of epidemiological interest} \label{appendix::steady_state_dist}

As a function of the FORI $\lambda(t)$ in the interval $\tau \in [0, t)$, we can recover distributions for various quantities of epidemiological interest at time $t$ using the results derived in \textcite{mehra2022hypnozoite}. In the case of the open network of infinite server queues described in Section \ref{sec::within_host_human}, we assume a functional form for $\lambda(t)$, wherein we interpret these quantities as probabilistic distributions at the within-host level. In the context of the hybrid transmission models constructed in Section \ref{sec::hybrid_models}, where we capture the time evolution of $\lambda(t)$ by coupling host and vector dynamics, these quantities are instead interpreted as population-level proportions \parencite{henry2020hybrid}. We adopt the same notation for both situations, with $N_s(t)$ denoting the marginal distribution for the number of hypnozoites/infections in state $s \in S$ at time $t$ on either the within-host or population-level.

\subsection{Size of the (non)-latent hypnozoite reservoir}
The expected size of the (non)-latent hypnozoite reservoir at time $t$
\begin{align}
    \EX[ N_{(N)L}(t) ] &= \nu \int^t_0 \lambda(\tau) p_{h,(N)L}(t-\tau) d\tau \label{exp_hyp}
\end{align}
and associated variance
\begin{align}
    \text{Var} [ N_{(N)L}(t)] &= \nu \int^\infty_0  \lambda(\tau) p_{h,(N)L}(t-\tau) \big[ 1 + 2 \nu p_{h,(N)L}(t-\tau) \big] d \tau. \label{var_hyp}
\end{align}
follow from Equations (39) and (74) of \textcite{mehra2022hypnozoite}. Complete distributions for $S_{(N)L}(t)$ can be recovered using Equations (75) and (76) of \textcite{mehra2022hypnozoite}. For short-latency strains ($k=0$), a constant FORI $\lambda(\tau) = \lambda^*$ yields the steady state distribution
\begin{align*}
    N^*_{NL} \sim \text{NegativeBinomial} \Big( \frac{\nu}{1+\nu}, \frac{\lambda^*}{\alpha + \mu} \Big),
\end{align*}
as shown in Equation (36) of \textcite{mehra2022hypnozoite}.

\subsection{Blood-stage infection status}
Equations (82) and (83) of \textcite{mehra2022hypnozoite} yield the complete population-level distribution for the number of co-circulating parasite broods. Here, we state the respective probabilities of carrying $n=0,1,2$ broods of co-circulating parasites at time $t$:
\begin{align}
    P(N_A(t) + N_P(t) = 0) &= \exp \bigg\{ -\int^t_0 \lambda(\tau) \frac{p_{p,A}(t-\tau) + \nu p_{h,A}(t-\tau)}{1 + \nu p_{h,A}(t-\tau)} d \tau \bigg\} \label{moi_0}\\
    P(N_A(t) + N_P(t) = 1) &=  \bigg( \int^t_0 \lambda(\tau) \frac{p_{p,A}(t-\tau) + \nu p_{h,A}(t-\tau)}{1 + \nu p_{h,A}(t-\tau)} d \tau \bigg) \notag \\
    & \qquad \times \exp \bigg\{ -\int^t_0 \lambda(\tau) \frac{p_{p,A}(t-\tau) + \nu p_{h,A}(t-\tau)}{1 + \nu p_{h,A}(t-\tau)} d \tau \bigg\} \label{moi_1}\\
    P(N_A(t) + N_P(t) = 2) &= \frac{1}{2} \bigg[  \bigg( \int^t_0 \lambda(\tau) \frac{p_{p,A}(t-\tau) + \nu p_{h,A}(t-\tau)}{1 + \nu p_{h,A}(t-\tau)} d\tau \bigg)^2 \notag \\
    & \qquad \qquad + 2 \int^t_0 \lambda(\tau) \frac{\nu p_{h,A}(t-\tau) \big(  p_{p,A}(t-\tau) + \nu p_{h,A}(t-\tau) \big) }{\big(1 + \nu p_{h,A}(t-\tau) \big)^3} d\tau \bigg] \notag \\
    & \qquad \times \exp \bigg\{ -\int^t_0 \lambda(\tau) \frac{p_{p,A}(t-\tau) + \nu p_{h,A}(t-\tau)}{1 + \nu p_{h,A}(t-\tau)} d \tau \bigg\}. \label{moi_2}
\end{align}

\subsection{Relapse rate conditional on blood-stage infection status}

The joint PGF for the size of the non-latent hypnozoite reservoir $N_{NL}(t)$ and the number of co-circulating parasite broods $M_I(t) := N_A(t) + N_P(t)$ can be written
\begin{align}
    \EX \big[ x^{NL(t)} y^{M_I(t)} ] &= G(z_1 = 1, \dots, z_k=1, z_{NL}=x, z_A=y, z_C=1, z_D=1, z_P=y)\\
    &= \exp \Bigg\{ -\int^t_0 \lambda(\tau) \Bigg[ 1 - \frac{1 + (y-1) p_{p,A}(t-\tau) }{1 + \nu( 1 - x) p_{h,NL}(t) + \nu(1-y) p_{h,A}(t)} \Bigg] d \tau \Bigg\} \label{rel_rate_unconditional}
\end{align}
where we have used Equation (\ref{vivax_multi_pgf_imm}).\\

The (unconditional) relapse rate, which is proportional to the expected size of the non-latent hypnozoite reservoir, can be written
\begin{align}
    r(t) &:= \alpha \EX[N_{NL}(t)] = \alpha \nu \int^t_0 \lambda(\tau)  p_{h,NL}(t-\tau) d \tau \label{rel_rate_moi_blood}
\end{align}
as in Equation (39) of \textcite{mehra2022hypnozoite}.\\

Conditional on the absence of blood-stage infection ($M_I(t)=0$), the relapse rate is given by 
\begin{align}
    r_0(t) &:= \alpha \EX[N_{NL}(t) | M_I(t) = 0] = \frac{\frac{\partial}{\partial x} \EX \big[ x^{NL(t)} y^{M_I(t)} ]|_{x=1, y=0}}{\EX \big[ x^{NL(t)} y^{M_I(t)} ]|_{x=1, y=0}}\\
    & = \alpha \int^t_{0} \lambda(\tau) \frac{ \nu p_{h,NL}(t-\tau) \big( 1 - p_{p, A}(t-\tau) \big)}{\big( 1 + \nu p_{h,A}(t-\tau) \big)^2} d \tau. \label{rel_rate_moi_0}
\end{align}
as in Equation (78) of \textcite{mehra2022hypnozoite}.\\

Likewise, by \textcite{xekalaki1987method}, we recover the relapse rate conditional on a single-brood blood-stage infection ($M_I(t)=1$),
\begin{align}
    r_1(t) &:= \alpha \EX[N_{NL}(t) | M_I(t) = 1] = \frac{\frac{\partial^2}{\partial x \partial y} \EX \big[ x^{NL(t)} y^{M_I(t)} ]|_{x=1, y=0}}{\frac{\partial}{\partial y}\EX \big[ x^{NL(t)} y^{M_I(t)} ]|_{x=1, y=0}} \notag \\
    & = \alpha \Bigg[ \frac{\int^t_0 \lambda(\tau) \frac{\nu p_{h,NL}(t-\tau) [ 2 \nu p_{h,A}(t-\tau) + p_{p,A}(t-\tau) ( 1 - \nu p_{h,A}(t-\tau)) ]}{ (1 + \nu p_{h,A}(t-\tau) )^3} d \tau }{\int^t_0 \lambda(\tau) \frac{\nu p_{h,A}(t-\tau) + p_{p,A}(t-\tau)}{ (1 + \nu p_{h,A}(t-\tau) )^2}  d \tau} \notag \\
    & \qquad \qquad + \int^t_{0} \lambda(\tau) \frac{ \nu p_{h,NL}(t-\tau) \big( 1 - p_{p, A}(t-\tau) \big)}{\big( 1 + \nu p_{h,A}(t-\tau) \big)^2} d \tau \Bigg]. \label{rel_rate_moi_1}
\end{align}
as well as a double-brood blood-stage infection ($M_I(t) = 2$)
\begin{tiny}
\begin{align}
    r_2(t) &:= \alpha \EX[N_{NL}(t) | M_I(t) = 2] = \frac{\frac{\partial^3}{\partial x \partial y^2} \EX \big[ x^{NL(t)} y^{M_I(t)} ]|_{x=1, y=0}}{\frac{\partial^2}{\partial y^2}\EX \big[ x^{NL(t)} y^{M_I(t)} ]|_{x=1, y=0}} \notag \\
    & = \alpha \Bigg[ \Bigg(\int^t_0 \lambda(\tau) \frac{2 \nu^2 p_{h,NL}(t-\tau) p_{h,A}(t-\tau) [3 \nu p_{h,A}(t-\tau) + p_{p,A}(t-\tau) ( 2 - \nu p_{h,A}(t-\tau)) ] }{(1 + \nu p_{h,A}(t-\tau))^4} d \tau \notag \\
    & \qquad \qquad \qquad + 2 \Big[ \int^t_0 \lambda(\tau) \frac{\nu p_{h,A}(t-\tau) + p_{p,A}(t-\tau)}{ (1 + \nu p_{h,A}(t-\tau) )^2}  d \tau \Big] \Big[ \int^t_0 \lambda(\tau) \frac{\nu p_{h,NL}(t-\tau) [ 2 \nu p_{h,A}(t-\tau) + p_{p,A}(t-\tau) ( 1 - \nu p_{h,A}(t-\tau)) ]}{ (1 + \nu p_{h,A}(t-\tau) )^3} d \tau \Big] \Bigg) \notag \\
    & \qquad \qquad \times \Bigg( \int^t_0 \lambda(\tau) \frac{2 \nu p_{h,A}(t-\tau) [\nu p_{h,A}(t-\tau) + p_{p,A}(t-\tau)}{(1 + \nu p_{h,A}(t-\tau))^3} d \tau + \Big[ \int^t_0 \lambda(\tau) \frac{\nu p_{h,A}(t-\tau) + p_{p,A}(t-\tau)}{ (1 + \nu p_{h,A}(t-\tau) )^2}  d \tau \Big]^2\Bigg)^{-1} \notag \\
    & \qquad \qquad + \int^t_{0} \lambda(\tau) \frac{ \nu p_{h,NL}(t-\tau) \big( 1 - p_{p, A}(t-\tau) \big)}{\big( 1 + \nu p_{h,A}(t-\tau) \big)^2} d \tau \Bigg]. \label{rel_rate_moi_2}
\end{align}
\end{tiny}

By the law of total expectation, we can recover the relapse rate conditional on a blood-stage infection comprising three or more parasite broods ($M_I(t)>2$):
\begin{align}
    r_{>2}(t) &:= \alpha \EX[N_{NL}(t) | M_I(t) > 2]\\
    & = \frac{r(t) - r_0(t) \cdot P(M_I(t) = 0) - r_1(t) \cdot P(M_I(t) = 1) - r_2(t) \cdot P(M_I(t) = 2) }{ 1 - P(M_I(t) = 0) - P(M_I(t) = 1) - P(M_I(t) = 2)}. \label{rel_rate_moi_3+}
\end{align}
using Equations (\ref{moi_0}) to (\ref{moi_2}) and (\ref{rel_rate_moi_blood}) to (\ref{rel_rate_moi_2}).

\subsection{Relative contribution of relapses to the infection burden}
At time $t$, each individual in the human population has no ongoing primary infections with probability
\begin{align*}
    P(N_P(t) = 0) = \exp \bigg\{ -\int^t_0 \lambda (\tau) p_{p,I}(t-\tau) d\tau \bigg\}
\end{align*}
and no ongoing relapses with probability
\begin{align*}
    P(N_A(t) = 0) = \exp \bigg\{ -\int^t_0 \lambda (\tau) \frac{\nu p_{h,A}(t-\tau)}{1 + \nu p_{h,A}(t-\tau)} d\tau \bigg\}
\end{align*}
as in Equations (40) and (41) of \textcite{mehra2022hypnozoite}. The relative contribution of relapses to the infection burden (that is, the proportion of blood-stage infections that encompass at least one relapse) follows readily from the quantities $P(N_P(t) = 0)$, $P(N_A(t) = 0)$ and $P(N_A(t) = N_P(t) = 0)$ (that is, the probability that each individual has neither a primary infection, nor a relapse, at time $t$).

\subsection{Distribution of immunity}
From the joint PGF given by Equation (\ref{vivax_multi_pgf_imm}), we recover the PGF governing the population-level distribution of immunity $I(t)$ at time $t$  
\begin{align*}
    \EX \big[ z^{N_I(t)} ] &= G(t, z_1=1, \dots, z_k=1, z_{NL}=1, z_A=1, z_D=1, z_C=1, z_I=z, z_P=1)\\
    &= \exp \bigg\{ - \int^t_0 a I_M(\tau) \bigg[ 1 - \frac{1 - (1-z) p_{p,I}(t-\tau)}{1 + \nu p_{h,I}(t-\tau) (1-z)} \bigg] d \tau \bigg\}.
\end{align*}

We can thus compute the expected immunity level
\begin{align}
    \EX[ N_I(t) ] =& \frac{\partial \EX \big[ z^{P_I(t)} ] }{\partial z} \Bigg|_{z=1} =  \int^t_0 \lambda(\tau) \big[ \nu p_{h,I}(t-\tau) + p_{p,I}(t-\tau) \big] dt \label{exp_imm}
\end{align}
in addition to the variance
\begin{align}
    \text{Var}[ N_I(t) ] =& \frac{\partial^2 \EX \big[ z^{N_I(t)} ] }{\partial z^2} \Bigg|_{z=1} + \frac{\partial \EX \big[ z^{N_I(t)} ] }{\partial z} \Bigg|_{z=1} - \bigg( \frac{\partial \EX \big[ z^{N_I(t)} ] }{\partial z} \Bigg|_{z=1} \bigg)^2 \notag \\
    =&  \int^t_0 \lambda (\tau) \big[ \nu p_{h,I}(t-\tau) + p_{p,I}(t-\tau) \big] \big[ 1 + 2 \nu  p_{h,I}(t-\tau) \big] dt. \label{var_imm}
\end{align}

The complete population-level distribution of immunity can be recovered using Faa di Bruno's formula and Leibiniz's integral rule, using a similar approach to \textcite{mehra2022hypnozoite} (see Equations (82) and (83) of \textcite{mehra2022hypnozoite}, which have an analogous functional form to that consided here).

\section{Expected vector dynamics} \label{appendix:expected_vector}
Here, we characterise the dynamics of the vector population, as a function of the probability of successful human-to-mosquito transmission $p_{h \to m}(t)$. The structure of the Markovian birth-death process governing the state of the vector population is described in Section \ref{sec::vector_dynamics}.\\

Denote by $p_{i, l, u}(t)$ the probability that the mosquito population comprises $i$ infected, $l$ latent and $u$ uninfected mosquitoes at time $t$. By the Kolmogorov forward differential equations, it follows that
\begin{align}
    \frac{d p_{i,l, u}}{dt} = & \underbrace{g \big[ -(i + u + l) p_{i,l,u}(t) + (i+1) p_{i+1, l, u}(t) + (u+1) p_{i, l, u+1}(t) + (l+1) p_{i, l+1, u}(t)]}_\text{mosquito death} \notag \\
    & + \underbrace{\omega(t) \big[ - (i + l + u) p_{i, l, u}(t) + (i + l + u - 1) p_{i, l, u-1}(t) \big]}_\text{mosquito birth} \notag \\
    & + \underbrace{\beta(t) p_{h \to m}(t) \big[ - u p_{i, l, u}(t) + (u + 1) p_{i, l-1, u+1}(t) \big]}_\text{human-to-mosquito transmission} \notag \\
    &+ \underbrace{\eta \big[ - l p_{i, l, u}(t) + (l + 1) p_{i-1, l+1,u}(t) \big]}_\text{sporogony complete}. \label{mos_chapman_kolmogorov}
\end{align}

Define the generating function
\begin{align*}
    F(z_i, z_l, z_u, t) = \sum^\infty_{j_i = 0} \sum^\infty_{j_l = 0} \sum^\infty_{j_u = 0} p_{j_i, j_l, j_u}(t) z_i^{j_i} z_l^{j_l} z_u^{j_j}.
\end{align*}

Using the ODEs given by Equation (\ref{mos_chapman_kolmogorov}), we recover a PDE for the generating function $F$
\begin{align}
    \frac{\partial F}{\partial t} =&  +  \big[ g(1-z_i) + \omega(t) z_i (z_u - 1) \big] \cdot \frac{\partial F}{\partial z_i} \notag \\
    & + \big[ g(1-z_l) + \omega(t) z_l (z_u - 1) + \eta (z_i - z_l) \big] \cdot \frac{\partial F}{\partial z_u} \notag \\
    & + \big[ g(1-z_u) + \omega(t) z_u (z_u - 1) + \beta(t) p_{h \to m}(t) (z_l - z_u) \big] \cdot \frac{\partial F}{\partial z_u}. \label{mos_GF_PDE}
\end{align}

Denote by $I_M(t)$, $L_M(t)$, $U_M(t)$ the expected number of infected, latent and uninfected mosquitoes respectively at time $t$, that is,
\begin{align*}
    I_M(t) = \frac{\partial F}{\partial z_i} \Big|_{z_i = z_l = z_u = 1} \qquad L_M(t) = \frac{\partial F}{\partial z_l} \Big|_{z_i = z_l = z_u = 1} \qquad U_M(t) = \frac{\partial F}{\partial z_u} \Big|_{z_i = z_l z_u = 1}.
\end{align*}

Then for $s \in \{i, l, u\}$, by differentiating the PDE given by Equation (\ref{mos_GF_PDE}) with respect to $z_s$ and evaluating the resultant expression at $z_i = z_l = z_u = 1$, we recover precisely the system of ODEs given by Equations (\ref{d_im}) to (\ref{d_um}) governing the time evolution of $I_M(t)$, $L_M(t)$, $U_M(t)$ as a function of the probability of human-to-mosquito transmission $p_{h \to m}(t)$ per bloodmeal. More detailed vector dynamics can be captured in an analogous manner.

\section{The existence of endemic equilibrium solutions} \label{appendix::endemic_eq}
For notational convenience, let
\begin{align*}
    A &:= \frac{\beta p_{m \to h}}{P_H} \int^\infty_0 \frac{1 - (1-p_{tb}) p_{p, I}(\tau)}{1 + \nu (1-p_{tb}) p_{h, I}(\tau)} - \frac{1 - (1-p_{tb}) p_{p, I}(\tau) - p_{p, A}(\tau)}{1 + \nu (1-p_{tb}) p_{h, I}(\tau) + \nu p_{h, A}(\tau)} d \tau\\
    C &:= \frac{\beta p_{m \to h}}{P_H} \int^\infty_0 1 - \frac{1 - (1-p_{tb}) p_{p, I}(\tau)}{1 + \nu (1-p_{tb}) p_{h, I}(\tau)} d \tau.
\end{align*}

From Equations (\ref{c_imm_exp_1}) and (\ref{c_imm_exp_2}), we observe that
\begin{align*}
    e^{-C I_M^*} &= \EX \big[ p_{tb}^{N_I(t)} \big] = \sum^\infty_{i_1=0} \dots \sum^\infty_{i_k=0} \sum^\infty_{i_{NL}=0} \sum^\infty_{j=0} \sum^\infty_{k=0} p_{tb}^k H^*_{i_1, \dots, i_k, i_{NL}, j, k} \leq 1\\
    e^{-( A + C) I_M^*} &= \EX \big[ p_{tb}^{N_I(t)} | N_A(t) + N_P(t) = 0 \big] \cdot P(N_A(t) + N_P(t) = 0)\\
    & = \sum^\infty_{i_1=0} \dots \sum^\infty_{i_k=0} \sum^\infty_{i_{NL}=0} \sum^\infty_{k=0} p_{tb}^k H^*_{i_1, \dots, i_k, i_{NL}, 0, k} \leq e^{-C I_M^*}
\end{align*}
from which we deduce that $A, C \geq 0$.\\

Recall from Section \ref{sec::steady_state} that the existence of an endemic equilibrium solution is equivalent to the existence of a non-trivial solution to Equations (\ref{im_sim_eq_2}) and (\ref{im_sim_eq_1}). Characterising endemic equilibria is thus equivalent to characterising non-trivial points of intersection $I_M^* \in (0, P_M]$ of
\begin{align*}
    F_1(I_M^*) &= \frac{g I_M^*}{\beta p_0 \big(\frac{P_M}{1 + g/\eta}- I_M^* \big)}\\
    F_2(I_M^*) &= e^{-C I_M^*} - e^{-(A+C)I_M^*}.
\end{align*}

We note that $F_2(I_M^*) \geq 0$ for all $I_M^* \geq 0$ since $A, C \geq 0$, while $F_1(I^*_M) \geq 0$ only if $I_M^* \leq \frac{P_M}{1 + g/\eta}$. We thus seek solutions in the domain $I_M^* \in (0, \frac{P_M}{1 + g/\eta} \big]$.\\

Observe that $F_1(0) = F_2(0) = 0$, corresponding to the disease-free equilibrium. To characterise endemic equilibria, we compute the derivatives
\begin{align*}
    F_1^{(n)}(I_M^*) &= n! \frac{g P_M}{\beta p_0 ( 1 + g/\eta)} \frac{1}{ \big( \frac{P_M}{1 + g/\eta} - I_M^* \big)^{n+1}}\\
    F_2^{(n)}(I_M^*) &= (-1)^n(C)^n e^{-C I_M^*} + (-1)^{n+1} (C+A)^n e^{-(C+A)I_M^*}
\end{align*}

For all $I_M^* \in [0, \frac{P_M}{1 + g/\eta}]$, $F'_1(I_M^*), F_1''(I_M^*) >0$, so we conclude that $F_1$ is monotonically increasing and convex, with $F_1 \to \infty$ as $I_M^* \to \frac{P_M}{1 + g/\eta}$.\\

On the other hand, observing that $F_2'$, $F_2''$ are continuous; $F_2'(0) > 0$, $F_2^{''}(0) < 0$ and
\begin{align*}
    &F_2'(M_0) = 0 \implies M_0 = \frac{1}{A} \log \Big(1 + \frac{A}{C} \Big)\\
    &F_2''(M_1) = 0 \implies M_1 = \frac{2}{A} \log \Big(1 + \frac{A}{C} \Big) > M_0,
\end{align*}
by the intermediate value theorem, we deduce that $F_2$ is
\begin{itemize}
    \item monotonically increasing, concave for all $I^*_M \in [0, M_0)$
    \item monotonically decreasing for all $I^*_M \in (M_0, \infty)$.
\end{itemize}

Setting $Q := F_1' - F_2'$, we note that:
\begin{itemize}
    \item $Q'(I_M^*) > 0$ for $I_M^* \in [0, M_0)$
    \item $Q(I_M^*) > 0$ for all $I_M^* \in [M_0, \frac{P_M}{1+g/\eta})$
\end{itemize}

We thus consider two cases:
\begin{itemize}
    \item \textbf{Case 1}: $Q(0) > 0$\\
    Then $Q(I_M^*) > 0$ for all $I_M^* \in [0, P_M] \implies (F_1 - F_2)$ is monotonically increasing for all $I_M^* \in [0, 1]$. Since $(F_1-F_2)(0) = 0$, there exists no non-trivial point of intersection.
    \item \textbf{Case 2}: $Q(0) < 0$\\
    Since $Q$ is monotonically increasing on the interval $[0, M_0)$ with $Q(0) < 0 < Q(M_0)$, by the intermediate value theorem, there exists $M_1 \in (0, M_0)$ such that $Q(I_M^*) \leq 0$ for $I_M^* \in [0, M_1]$ and $Q(I_M^*) > 0$ for $I_M^* \in (M_1, \infty)$. Noting that $(F_1 - F_2)(0) = 0$, it follows that  $(F_1 - F_2)$ is montonically decreasing and negative on the interval $(0, M_1)$; but monotonically increasing on the interval $(M_1, \frac{P_M}{1+g/\eta})$. Since $(F_1 - F_2)(I_M^*) \to + \infty$ as $I_M^* \to \frac{P_M}{1 + g/\eta}$, by the intermediate value theorem, there exists a unique non-trivial root of $(F_1 - F_2)$ in the interval $(M_1, \frac{P_M}{1 + g/\eta})$.
\end{itemize}

Therefore, $F_1$ and $F_2$ have at most one non-zero point of intersection if and only if
\begin{align*}
    Q(0) < 0 \iff R_0^2 := \frac{A \cdot \beta p_0 P_M}{g (1 + g/\eta )} > 1.
\end{align*}
As such, $R_0 > 1$ is a sufficient and necessary condition for the existence of an endemic equilibrium; when $R_0 > 1$, the endemic equilibrium solution is unique.

\newpage

\printbibliography

\end{document}